\documentclass[11pt,letterpaper]{article}

\usepackage[margin=1in]{geometry}
\usepackage[T1]{fontenc}
\usepackage[utf8]{inputenc}
\usepackage{newtxtext}
\usepackage{helvet}
\usepackage{courier}
\usepackage{amsmath}
\usepackage{graphicx}
\usepackage{array}
\usepackage{tabularx}
\usepackage{booktabs}
\newcolumntype{L}[1]{>{\raggedright\arraybackslash}p{#1}}
\newcolumntype{Y}{>{\raggedright\arraybackslash}X}
\usepackage{caption}
\usepackage{natbib}
\usepackage[hyphens]{url}
\usepackage{microtype}
\usepackage[hidelinks]{hyperref}

\title{\bfseries VibeMemBench: Evaluating Memory Systems for Coding Agents\\
on Real Repository Coding Tasks}

\author{%
  Liyang Fan$^{1,4}$ \quad Yingcheng Shi $^{2}$ \quad Yongbin Li$^{2,\dagger}$ \quad
  Chenghao Sun$^{1}$ \quad Xin Chen$^{3}$\\
  Xander Xu$^{3}$ \quad Hu Wei$^{3}$ \quad Shiwen Ni$^{4,\dagger}$ \quad
  Min Yang$^{1,4,\dagger}$ \quad Jieping Ye$^{2}$\\[0.7em]
  \normalsize  \qquad $^{1}$Shenzhen Institutes of Advanced Technology, Chinese Academy of Sciences \\ \qquad $^{2}$Alibaba Token Hub, Alibaba Group \qquad $^{3}$Alibaba Group \qquad $^{4}$SUAT \\[0.5em]
}

\date{}

\begin{document}

\maketitle

\begingroup
\renewcommand{\thefootnote}{}
\footnotetext{$^{\dagger}$Corresponding authors: shuide.lyb@alibaba-inc.com, min.yang@siat.ac.cn.\\
Code available at: \url{https://github.com/AlibabaResearch/DAMO-ConvAI/tree/main/VibeMemBench}}
\endgroup

\begin{abstract}
Coding agents operate on real repository coding tasks, and persistent memory systems promise to reuse experience across tasks.  Yet existing evaluations do not show whether those systems improve executable repository work.  Repository benchmarks test code changes but do not isolate memory, while memory benchmarks score recall without measuring downstream coding outcomes.  We introduce VibeMemBench, a benchmark for evaluating memory systems on 111 coding targets from 90 SWE-rebench V2 repositories and 3,634 history trajectories from the target repositories.  The targets follow the SWE benchmark style and cover bug fixes, feature requests, interface changes, and configuration work.  An agent edits each target codebase under a declared memory condition.  Executable tests decide task resolution.  Each target is retained only when injected history experience improves its executable outcome in a reference setting, so every target carries a prior experience whose usefulness is verified by execution in that setting.  The frozen verified experience is then transferred to five held-out solvers.  Direct injection raises observed task resolution on four of them by 1.1 to 4.5 percentage points while lowering agent steps on all five.  Yet when four existing memory systems must construct and retrieve experience from the same history, eleven of twelve solver and system pairings fail to exceed the matched memory-off baseline.  VibeMemBench exposes the gap between the useful experience that repository history holds and the experience existing memory systems deliver for repository coding tasks.
\end{abstract}

\section{Introduction}

Persistent memory in coding agents presents a measurement problem.  Executable
tests give coding tasks a grounded outcome, while useful memory remains latent.
No instance-level label says which earlier observation a system ought to
write, retain, or retrieve, and a gold memory label would
answer a different question from whether memory improves work.

\begin{figure}[!t]
\centering
\includegraphics[width=\linewidth]{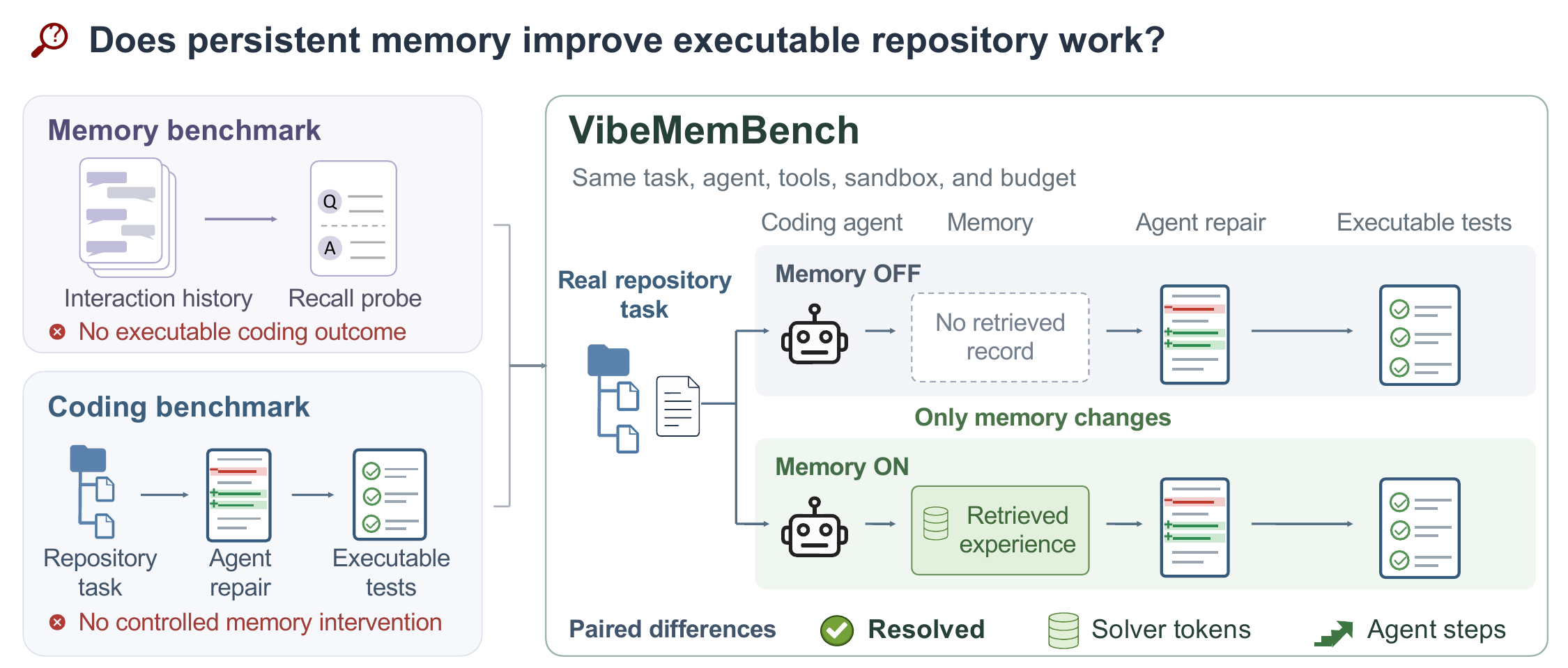}
\caption{From partial benchmark signals to a controlled memory comparison.
Left: memory benchmarks probe recall without an executable coding outcome,
while repository benchmarks test repairs without isolating a memory
intervention.  Right: VibeMemBench pairs two runs of the same repository task
under a fixed agent, tools, sandbox, and budget.  Memory off supplies no
retrieved record; the depicted memory-on condition supplies a retrieved
experience.  Executable tests determine task resolution.  The comparison
reports paired differences in Resolved, solver tokens, and agent steps.}
\label{fig:motivation}
\end{figure}

Existing evaluation families observe only one side of this problem.
Memory benchmarks probe facts from interaction histories
\citep{maharana2024locomo,wu2025longmemeval}, but recall alone does not
establish whether an experience helps a coding agent repair a repository.
Repository benchmarks measure executable task outcomes
\citep{jimenez2024swebench}, but do not isolate a persistent memory
intervention.  The left panels of Figure~\ref{fig:motivation} show these
two partial signals.

The right panel turns memory evaluation into a matched comparison of
repository work.  Both runs share the task, agent, tools, sandbox, and
budget.  Only the declared memory condition changes.  Executable tests
measure whether the supplied experience improves task resolution rather
than whether it matches a gold retrieval label.  We report Resolved with
solver token consumption and agent steps.  The latter two measure solver
use, not latency or total memory-system resource consumption.

VibeMemBench builds on SWE-rebench V2, a large-scale corpus of
language-agnostic software engineering tasks with executable validation
\citep{badertdinov2026swerebenchv2languageagnosticswe}.  A four-stage
construction pipeline builds the targets.  We name it SIEVE for its
filtering stages, not as an acronym.  It
screens history from the same repository by patch patterns and keeps
only targets whose injected experience shows uplift verified by execution.  The benchmark comprises 111 targets across 90
repositories and 3,634 completed history
trajectories.  Each released target retains provenance, its test
command, and its execution result, so the task outcome is auditable
without a gold retrieval label.

The paper makes three contributions.
\begin{itemize}
    \item We construct VibeMemBench, a benchmark whose
    3,634 trajectories from the target repositories supply a persistent memory
    substrate and whose executable tests determine task resolution.  SIEVE
    controls leakage and verifies that a useful experience exists for
    every retained target.
    \item We formulate a matched intervention protocol for agent components
    whose intermediate choices lack an oracle.  It changes only the
    declared memory condition and separates
    frozen verified experience transfer from retrieval
    by existing memory systems under an auditable reporting contract.
    \item We report a gap between availability and use.  The frozen verified experience yields small positive aggregate
    Resolved differences for four of five transfer solvers, while retrieval
    by existing memory systems fails to exceed the matched memory-off
    baseline in eleven of twelve solver and system pairings.  Failure
    attribution locates the dominant break at the form of the supplied
    record, which a strip ablation traces to transcript volume rather than
    instruction semantics.  The value of a record is further conditioned on
    solver headroom rather than intrinsic to it, so the same record helps a
    weaker solver and harms a stronger one.
\end{itemize}

\section{Related Work}

\paragraph{Coding benchmarks.}
SWE-bench resolves GitHub issues through executable tests
\citep{jimenez2024swebench}.  RepoBench, RepoCoder, and Hierarchical Context
Pruning study repository context, retrieval, and cross-file completion
\citep{liu2024repobench,zhang2023repocoder,zhang2025hierarchicalcontextpruning}.
GitTaskBench and DSCodeBench make repository artifacts, task criteria, and
executable validation explicit \citep{ni2026gittaskbench,ouyang2026dscodebench}.
Talk2Code evaluates multi-turn code generation with dual-track code and
interaction criteria \citep{yang2026talk2code}, and SWE-Bench-CL orders
software engineering tasks into continual learning sequences
\citep{joshi2025swebenchcl}.  None of these benchmarks isolates a matched
persistent memory intervention across coding work.

\paragraph{Long-horizon agent benchmarks.}
DMT-RoleBench measures dynamic multi-turn role-playing
\citep{yuan2025dmtrolebench}, RealWebAssist studies long-horizon web assistance
with real users over time \citep{ye2026realwebassist}, and ProBench evaluates GUI agents
with process information \citep{yang2026probench}.  Their primary evaluation signal is
dialogue, web, or GUI behavior rather than a matched memory-on versus
memory-off effect on executable repository repair.

\paragraph{Memory benchmarks.}
LoCoMo and LongMemEval evaluate long
multi-session histories through question answering, temporal reasoning, and
knowledge updates \citep{maharana2024locomo,wu2025longmemeval}, and MemBench
broadens this setting with factual and reflective memory under multiple
interaction scenarios \citep{tan2025membench}.  Their primary signal is a
queried answer over stored history.  They do not measure whether latent useful
memory changes executable coding success.

\paragraph{Memory systems as evaluated components.}
Persistent memory systems supply write, update, retention, and retrieval policies.  MemGPT,
MemoryBank, MEMORYLLM, MemoryART, and LightMem develop tiered context
management, long-term personal stores, self-updatable latent memory, and
lightweight record extraction
\citep{packer2023memgpt,zhong2024memorybank,wang2024memoryllm,dai2026memoryart,fang2025lightmem}.
VibeMemBench evaluates Mem0, SimpleMem, MemoryOS, and A-MEM
\citep{chhikara2025mem0,liu2026simplemem,kang2025memoryos,xu2025amem} as
four memory-on components under one shared executable protocol for
repository tasks, filling the gap these families leave.

\section{VibeMemBench}

\begin{figure}[t]
\centering
\includegraphics[width=\linewidth]{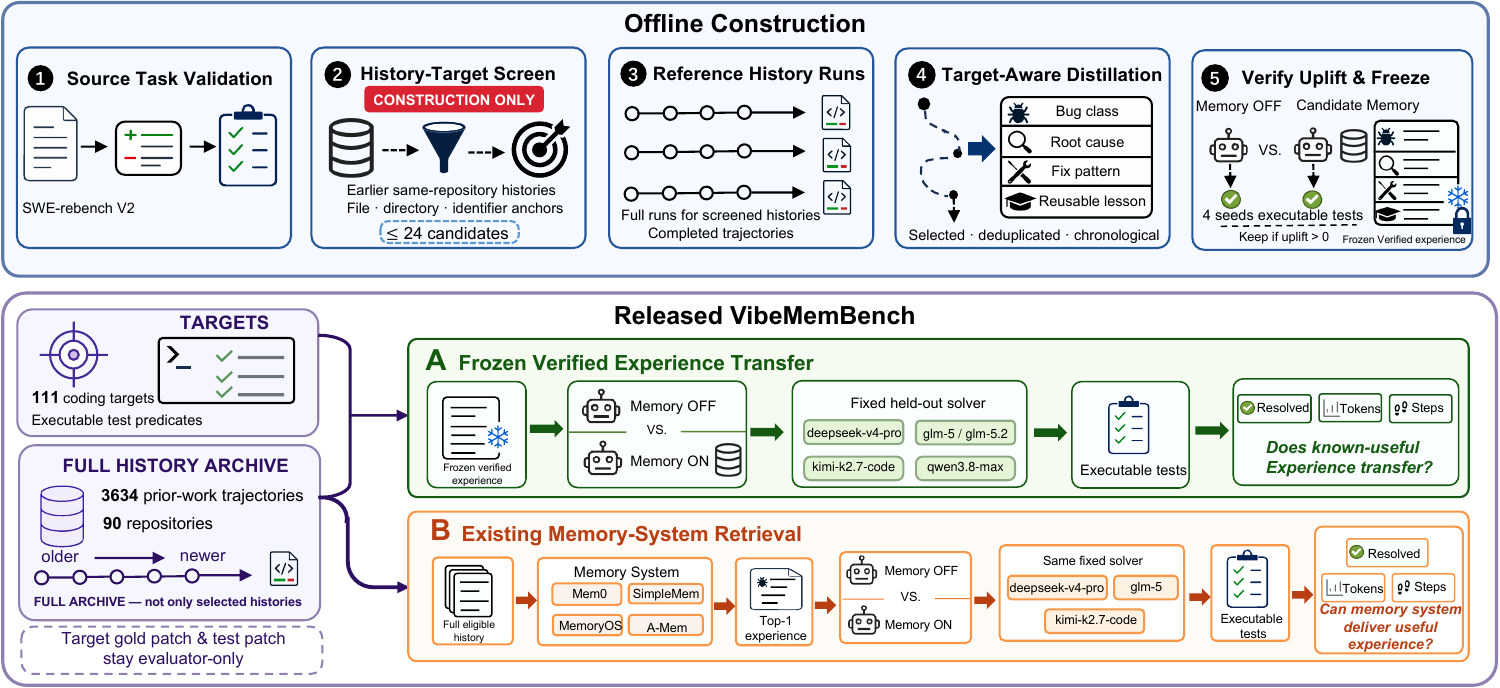}
\caption{VibeMemBench construction and evaluation overview.  The upper lane
shows offline construction.  The lower panel shows the released target-history
archive and the two evaluation layers.}
\label{fig:overview}
\end{figure}

The benchmark resolves a missing evaluation signal.  It makes the downstream
coding outcome executable while leaving the selection of useful memory latent.
One target is a repository state, a coding instruction, prior
trajectory observations, and an executable test predicate, and a run
succeeds when it passes.  The release preserves the command, environment, and result that
reproduce each run, and Figure~\ref{fig:overview} summarizes the
construction stages and both evaluation layers.

\subsection{Benchmark Profile}

\begin{figure}[t]
\centering
\includegraphics[width=\linewidth]{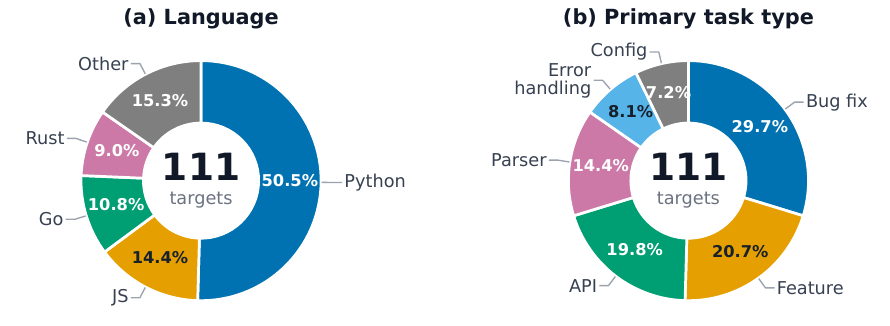}
\caption{Language (left) and primary task type (right) distributions of the 111 targets.
Other groups TypeScript, Swift, Kotlin, Java, and PHP.  The task type
chart merges the eleven audited types into six.  API adds behavior change
and compatibility fix, and Config adds refactor, test infrastructure,
and documentation change.  Appendix~\ref{app:task-type} gives all eleven
counts.}
\label{fig:target-profile}
\end{figure}

The left chart in Figure~\ref{fig:target-profile} reports the language
distribution across nine languages.  The distribution concentrates on
Python and three mainstream languages, so an aggregate score chiefly
reflects these languages.

The retained targets are repository-level coding tasks instantiated through a
SWE-style executable repair protocol, not a bug-repair slice.  A
construction audit, using artifacts that never reach the solver, assigns
each target one primary task type in the right chart of
Figure~\ref{fig:target-profile}.  Bug fixes account for under a third of
the labels, and feature, interface, configuration, and test
infrastructure work covers the rest.

\begin{figure}[t]
\centering
\includegraphics[width=0.78\linewidth]{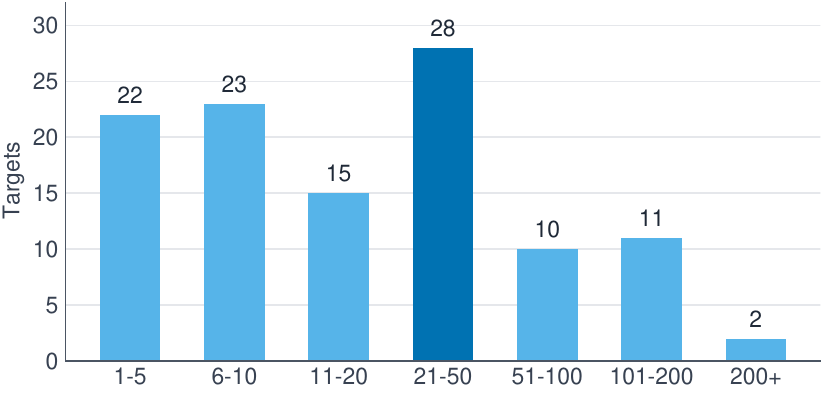}
\caption{History availability per target.  Bars count earlier trajectories
from the same repository.  Bins sum to 111 targets.}
\label{fig:history-profile}
\end{figure}

The 111 targets span 90 repositories after deduplication by history
directory and draw on 3,634 completed history trajectories.
Figure~\ref{fig:history-profile} reports the eligible history visible to each
target.  Targets from the same repository can share history trajectories, so the bars
describe target-level visibility.

\subsection{SIEVE Construction}

SIEVE constructs history-target pairs under leakage control and turns
history into target-aware experience.  Its stages are source validation,
patch-pattern screening, history execution with experience distillation, and
retention by uplift.

The first two stages control what enters the benchmark.  Validation runs the
declared tests with the source gold patch and discards every
instance whose gold patch fails.  The second stage is an oracle-assisted offline
patch-pattern screen.  It admits only earlier instances
from the same repository and excludes every target identifier, so no target enters its own
history pool.  A deterministic score over shared files,
directories, and cleaned identifiers shortlists candidates, and an LLM
judges each repair pattern with structured evidence.  The target gold
patch enters only this offline screen, the target test patch remains
evaluator only, and neither artifact reaches experience construction,
runtime memory retrieval, or the target solver context.
Appendix~\ref{app:source-fields} lists every inherited source field with
its visibility rule.

The third stage turns screened history into injectable experience.  SIEVE
executes each selected history task and keeps the completed trajectory.
Target-aware experience construction then distills one experience per pair
from the target problem statement, the history problem statement, a
compressed history trajectory, and the history final patch.  The target
statement selects which historical lesson to emphasize, and every concrete
claim remains anchored in the historical trajectory or patch.  The
distillation never sees the target gold patch, test patch, or test
outcome.  Each experience records a bug class, root cause, fix pattern,
and reusable lesson.

History execution uses one fixed reference MiniSWEAgent configuration
\citep{yang2024sweagent}, and this configuration, the scoring rule, the
shortlist defaults, and the gating policy appear in
Appendix~\ref{app:construction}.

\subsection{Target Selection and Frozen Verified Experience}

The benchmark uses a reference setting to establish that a useful prior
experience is available for a target.  A target-seed run is resolved when
the target's declared test predicate passes, and the resolved share of a
condition averages this binary outcome over its four seeds.  Let
$\mathcal{K}_t$ be the finite set of candidate-memory combinations that the
preceding SIEVE stages construct for target $t$, with the reserved index
$0\notin\mathcal{K}_t$ denoting the memory-off condition.  For
$k\in\mathcal{K}_t$, let $\widehat{p}^{\mathrm{ref}}_{t,k}$
be the resolved share for target $t$ after injecting
combination $k$, and let $\widehat{p}^{\mathrm{ref}}_{t,0}$ be the memory-off
share.  The uplift in the reference setting is
\begin{equation}
g_{t,k}=\widehat{p}^{\mathrm{ref}}_{t,k}-\widehat{p}^{\mathrm{ref}}_{t,0}.
\end{equation}
For every retained target, the frozen verified experience is the
combination $k_t^\star=\arg\max_{k\in\mathcal{K}_t} g_{t,k}$.  Retention requires
$\widehat{p}^{\mathrm{ref}}_{t,0}<1$ and $g_{t,k_t^\star}>0$.  This
rule is the fourth SIEVE stage.  Selection
uses deepseek-v4-flash \citep{deepseek2026models} and MiniSWEAgent with four
seeds per condition.  The argmax ranges over the shortlist
$\mathcal{K}_t$ rather than the full history pool, so $k_t^\star$
certifies that the history contains at least one useful experience for
the target, not that it is the most useful.

This reference comparison is a selection audit, reported apart from the
transfer results.  Its uplift is specific to the selected targets under
the reference solver, and the seed-level argmax lets seed noise inflate
it.  Cross-model transfer reuses the frozen verified experience and
reselects no memory combination for the evaluation model, so the
transfer estimates inherit the selected population but not the argmax
optimism.

\subsection{Memory Intervention}

Let $\mathcal{T}$ be the fixed set of coding targets, $\mathcal{S}$ the
set of evaluation seeds, and $\mathcal{P}=\mathcal{T}\times\mathcal{S}$
the paired evaluation set.  Let
$A$ denote the fixed agent configuration, including the model, prompt, tools,
sandbox image, context budget, resource limits, task order, and aggregation
rule, and let $m\in\{\mathrm{off},\mathrm{on}\}$ denote the
memory condition.  For each $(t,s)\in\mathcal{P}$, VibeMemBench compares
$A(t,s,\mathrm{off})$ and $A(t,s,\mathrm{on})$, and $m$ may change
only the declared persistence and retrieval mechanism.

The memory-on specification declares the write policy, the retrieval
trigger, the context insertion rule, and the record truncation rule, and
the memory-off control replaces those records under the same fixed
context budget.  Appendix~\ref{app:contract} states the full condition
contract and the manifest fields that every reported run fills.

\subsection{Evaluation Layers}

The first layer transfers the frozen verified experience that the
deepseek-v4-flash reference setting identifies for each retained
target.  The main comparison holds that experience fixed and evaluates
deepseek-v4-pro \citep{deepseek2026models}, glm-5
\citep{glm2026glm5}, glm-5.2 \citep{zai2026glm52docs}, kimi-k2.7-code
\citep{kimi2025k2,moonshot2026kimi27codedocs}, and qwen3.8-max
\citep{qwen2026qwen38max} with four seeds per
condition, testing whether the experience transfers across solver
models.  An irrelevant-memory control keeps the injection rule and
every other setting fixed but injects one fixed distractor passage of
everyday sentences with no software content.  The control and the
memory-system layer report deepseek-v4-pro, glm-5, and kimi-k2.7-code.
glm-5.2 and qwen3.8-max enter as transfer-only solvers.

The second layer evaluates retrieval by existing memory systems.  Mem0,
SimpleMem, MemoryOS, and A-MEM
\citep{chhikara2025mem0,liu2026simplemem,kang2025memoryos,xu2025amem} ingest
the full eligible prior-work history available to a target.
After ingestion and organization, a declared target query selects one
top-ranked experience, which the protocol injects into MiniSWEAgent.
This condition measures whether existing memory systems supply experiences that
improve downstream execution, and it is not online closed-loop memory, because
retrieval occurs before the target solver trajectory rather than during it.

For the existing memory systems, deepseek-v4-flash
\citep{deepseek2026models} extracts memory records from trajectories, Qwen3-Embedding-4B
\citep{zhang2025qwen3embedding} embeds retrieval queries and records, and
the release manifest records the retrieval configuration and the
target query template.

\section{Experiments}

Every reported comparison fixes the agent configuration $A$ of the
intervention protocol and varies only the memory condition, so an
observed difference is attributable to memory rather than to a changed agent
setup.

\subsection{Metrics}

The primary metric is task resolution.  We report \emph{Resolved}, the
share of target-seed runs whose declared test predicate passes, as
defined in the target selection stage.  For
$(t,s)\in\mathcal{P}$ and $m\in\{\mathrm{off},\mathrm{on}\}$, let
$y_{t,s}^{(m)}$ equal one when condition $m$ resolves target $t$ at seed $s$ by
passing the declared test predicate and zero otherwise.  The resolved
share $\widehat{p}_{m}$ for condition $m$ averages $y_{t,s}^{(m)}$ over
$\mathcal{P}$, and the paired memory effect is
$\widehat{\Delta}=\widehat{p}_{\mathrm{on}}-\widehat{p}_{\mathrm{off}}$.
The uncertainty unit is the target, so the reported intervals resample
the 111 targets with their four seeds intact 10,000 times and give
percentile bounds for $\widehat{\Delta}$.  Tabulated percentage-point
changes remain descriptive.

Every result row also reports solver resource use.  For each $(t,s)$ and
$m$, $u_{t,s}^{(m)}$ counts solver input and output tokens and
$h_{t,s}^{(m)}$ counts agent steps under one manifest-defined rule
shared by every condition.  The tables report the per-run means
$\widehat{u}_m$ and $\widehat{h}_m$ over $\mathcal{P}$.  Both measure
solver use and interaction length, not latency, and the protocol does
not record memory-side token consumption.  Wall-clock time and API price
vary with provider and pricing policy, so they do not enter the
comparison across systems.

\begin{table}[t]
\centering
{\footnotesize
\setlength{\tabcolsep}{3pt}
\begin{tabular*}{\linewidth}{@{\extracolsep{\fill}}lrrrrrr@{}}
\toprule
Condition & Resolved & 95\% CI & 4/4 & Input / run & Output / run & Steps / run \\
\midrule
\multicolumn{7}{@{}l}{\emph{deepseek-v4-pro}}\\
Memory off & 67.1\% & --- & 49 & 1.67M & 19.1k & 50.3 \\
Frozen verified experience & 67.1\% (+0.0) & $[-5.63,+5.86]$ & \textbf{51} (+2) & \textbf{1.49M} ($-$0.18M) & \textbf{18.9k} ($-$0.2k) & \textbf{47.6} ($-$2.7) \\
Irrelevant-memory control & 63.7\% ($-$3.4) & $[-8.11,+1.35]$ & 46 ($-$3) & 1.51M ($-$0.16M) & 18.1k ($-$1.0k) & 47.1 ($-$3.2) \\
\midrule
\multicolumn{7}{@{}l}{\emph{glm-5}}\\
Memory off & 55.0\% & --- & \textbf{35} & 2.45M & 15.4k & 77.1 \\
Frozen verified experience & \textbf{57.4\%} (+2.4) & $[-2.25,+7.21]$ & 34 ($-$1) & \textbf{1.93M} ($-$0.52M) & \textbf{13.2k} ($-$2.2k) & \textbf{66.2} ($-$10.9) \\
Irrelevant-memory control & 54.3\% ($-$0.7) & $[-5.63,+4.05]$ & 36 (+1) & 2.13M ($-$0.32M) & 14.5k ($-$0.9k) & 69.3 ($-$7.8) \\
\midrule
\multicolumn{7}{@{}l}{\emph{glm-5.2}}\\
Memory off & 76.8\% & --- & 67 & 4.06M & 45.9k & 68.2 \\
Frozen verified experience & \textbf{80.4\%} (+3.6) & $[-0.90,+8.11]$ & \textbf{74} (+7) & \textbf{3.40M} ($-$0.67M) & \textbf{39.3k} ($-$6.6k) & \textbf{61.0} ($-$7.2) \\
\midrule
\multicolumn{7}{@{}l}{\emph{kimi-k2.7-code}}\\
Memory off & 69.8\% & --- & 56 & \textbf{1.56M} & \textbf{16.2k} & 56.0 \\
Frozen verified experience & \textbf{74.3\%} (+4.5) & $[-0.23,+9.46]$ & \textbf{62} (+6) & 1.57M (+0.01M) & 16.4k (+0.2k) & \textbf{53.8} ($-$2.2) \\
Irrelevant-memory control & 69.1\% ($-$0.7) & $[-5.18,+3.83]$ & 57 (+1) & 1.62M (+0.06M) & 16.6k (+0.4k) & 55.9 ($-$0.1) \\
\midrule
\multicolumn{7}{@{}l}{\emph{qwen3.8-max}}\\
Memory off & 79.1\% & --- & 68 & 4.03M & \textbf{55.2k} & 52.9 \\
Frozen verified experience & \textbf{80.2\%} (+1.1) & $[-2.93,+5.41]$ & \textbf{70} (+2) & \textbf{4.00M} ($-$0.03M) & 60.3k (+5.1k) & \textbf{50.6} ($-$2.3) \\
\bottomrule
\end{tabular*}
}
\caption{Frozen verified experience transfer on the 111 outcome-selected targets.
Italic rows group the five transfer solvers.  Resolved is the share over 444
target-seed runs whose declared tests pass.  The
95\% CI column gives percentile bootstrap bounds in percentage points on the
paired Resolved difference from the memory-off row.  The
4/4 column counts targets that pass in every seed.  Input / run and
Output / run are solver-side mean input and output tokens per run, and
Steps / run is the mean count of declared agent steps.  Parentheses give
changes from the memory-off row, Resolved changes in percentage points and
other changes in displayed units.  Bold marks the better value
between the memory-off and frozen verified experience rows.  The
glm-5.2 and qwen3.8-max blocks have no irrelevant-memory control.}
\label{tab:transfer-results}
\end{table}

\begin{table}[t]
\centering
{\footnotesize
\setlength{\tabcolsep}{3pt}
\begin{tabular*}{\linewidth}{@{\extracolsep{\fill}}llrrrrr@{}}
\toprule
Solver & System & Resolved & 95\% CI & Input / run & Output / run & Steps / run \\
\midrule
deepseek-v4-pro & Mem0 & 62.8\% ($-$4.3) & $[-9.68,+1.35]$ & 1.75M (+0.08M) & 20.3k (+1.2k) & 51.2 (+0.9) \\
deepseek-v4-pro & SimpleMem & 66.9\% ($-$0.2) & $[-4.95,+4.50]$ & 1.54M ($-$0.13M) & 18.8k ($-$0.3k) & 48.7 ($-$1.6) \\
deepseek-v4-pro & MemoryOS & 64.6\% ($-$2.5) & $[-8.11,+3.15]$ & 1.70M (+0.03M) & 19.4k (+0.3k) & 50.9 (+0.6) \\
deepseek-v4-pro & A-MEM & 63.1\% ($-$4.0) & $[-10.14,+2.03]$ & 1.71M (+0.04M) & 20.0k (+0.9k) & 50.0 ($-$0.3) \\
\midrule
glm-5 & Mem0 & 49.5\% ($-$5.5) & $[-10.59,-0.45]$ & 2.17M ($-$0.28M) & 14.7k ($-$0.7k) & 71.7 ($-$5.4) \\
glm-5 & SimpleMem & 53.2\% ($-$1.8) & $[-6.76,+2.93]$ & 2.02M ($-$0.43M) & 13.3k ($-$2.1k) & 70.2 ($-$6.9) \\
glm-5 & MemoryOS & 57.0\% (+2.0) & $[-3.38,+7.43]$ & 1.96M ($-$0.49M) & 13.3k ($-$2.1k) & 67.2 ($-$9.9) \\
glm-5 & A-MEM & 52.5\% ($-$2.5) & $[-8.56,+3.38]$ & 2.01M ($-$0.44M) & 14.3k ($-$1.1k) & 66.5 ($-$10.6) \\
\midrule
kimi-k2.7-code & Mem0 & 68.0\% ($-$1.8) & $[-6.53,+2.70]$ & 1.61M (+0.05M) & 17.6k (+1.4k) & 57.0 (+1.0) \\
kimi-k2.7-code & SimpleMem & 69.6\% ($-$0.2) & $[-4.95,+4.50]$ & 1.62M (+0.06M) & 17.2k (+1.0k) & 56.1 (+0.1) \\
kimi-k2.7-code & MemoryOS & 66.0\% ($-$3.8) & $[-8.56,+1.13]$ & 1.60M (+0.04M) & 16.7k (+0.5k) & 53.9 ($-$2.1) \\
kimi-k2.7-code & A-MEM & 65.8\% ($-$4.0) & $[-9.68,+1.35]$ & 1.66M (+0.10M) & 16.4k (+0.2k) & 55.0 ($-$1.0) \\
\bottomrule
\end{tabular*}
}
\caption{Existing memory systems in the retrieval condition.
Resolved, input
tokens, output tokens, and steps use the same definitions as
Table~\ref{tab:transfer-results}.  Parentheses give changes from the
same-solver Memory off row in Table~\ref{tab:transfer-results}.  The 95\% CI
gives percentile bootstrap bounds on that paired
change.  Resolved
changes are percentage points and other changes use displayed units.}
\label{tab:retrieval-results}
\end{table}

\subsection{Results}

Three regularities organize Table~\ref{tab:transfer-results}.  The
frozen verified experience raises Resolved on four of the five transfer
solvers and leaves deepseek-v4-pro unchanged, so the aggregate transfer
effect is positive but not uniform.  All five bootstrap
intervals cross zero, so at 111 targets the observed gains of 1.1 to 4.5
points remain directional evidence rather than effects separable from
zero.  The irrelevant-memory control stays 0.7 to 3.4 points below the
memory-off baseline on each of the three solvers that run it, so
injected text alone reproduces none of
the gains.  The two largest gains add six and seven 4/4 targets, so
they are not confined to single-seed flips.  Injection also
lowers agent steps for every solver and input tokens for four of the
five, shortening the interaction rather than adding context cost.

The five gains do not order by memory-off strength.  glm-5 gains 2.4
points from the lowest baseline at 55.0 percent, deepseek-v4-pro gains
nothing at 67.1, kimi-k2.7-code gains the most at 4.5 from 69.8, glm-5.2
gains 3.6 from 76.8, and
qwen3.8-max gains 1.1 from the highest baseline at 79.1.  Aggregate
solver strength therefore does not predict the transfer effect, and the
mechanism developed below works at the level of the single target rather
than the solver.

Table~\ref{tab:retrieval-results} reports the retrieval condition for the
four existing memory systems on three of the five transfer solvers.  Eleven of the
twelve solver and system pairings stay at or below the matched memory-off
baseline, no system interval lies above zero, and the glm-5 and Mem0
interval lies entirely below it.  The single exception, MemoryOS on
glm-5, gains 2.0 points and
still trails the frozen verified experience.  This
nearly uniform shortfall separates the availability of a selected useful
experience from the ability of existing memory systems to supply useful
solver context.  All four systems cut glm-5 tokens and steps markedly,
yet only one converts the shorter interaction into a resolution gain,
so a shorter trajectory is not by itself evidence of useful memory.

The shortfall has a structural source in solver headroom.  Across the
1,332 retrieval pairings of target, solver, and system, the net effect is
a loss of
127 resolved seeds.  In 42.0 percent of those pairings a memory-off ceiling
of four resolved seeds out of four leaves an injected record able only to
lose seeds.  Every solver and condition combination gains seeds below the
ceiling and loses seeds at it.  The frozen verified experience offsets its
ceiling losses on all five solvers, and among the systems only the
largest gain does.  Appendix~\ref{app:headroom} stratifies all twenty
combinations.
Below the ceiling, existing memory systems help too little, and uniform
top-1 injection is not a neutral default.  A wider retrieval budget does
not repair the shortfall.  In Appendix~\ref{app:dose} a Mem0 dose curve on
deepseek-v4-pro stays below the memory-off baseline at every budget from
one to five records.  This matches the rare ranking miss that a wider
budget could mend.

The single positive pairing follows the same headroom mechanism rather
than a better record.  MemoryOS on glm-5 lifts 29 targets, with the seed
gains concentrated where glm-5 starts at zero or one resolved seed.  Those targets
receive raw transcripts of the form that dominates failure attribution,
and the same records help the stronger solvers less.  The pairing
interval crosses zero, so the gain reflects the headroom of the weakest
solver rather than a superior record.  Appendix~\ref{app:memoryos}
decomposes the pairing by memory-off bucket.

\subsection{Case Analysis}

Case analysis is predeclared before trace inspection.  Three matched
contrasts fix a solver across systems, fix a system across solvers on
the same injected record, or pair the frozen verified experience with
its memory-off run.  A target enters only when the two conditions differ
by at least two of four seeds.  Retrieval occurs before the target
trajectory, so a system injects identical text into every solver for the
same target.  A system contrast then isolates the supplied record, and a
solver contrast isolates its use.

Failure attribution makes the system contrast mechanical.  For each
failing pairing defined in Table~\ref{tab:failure-attribution}, the
diagnostic compares the top-1 injected record against the frozen
verified experience and assigns the first broken stage.  An anchor is a
shared file, directory, or identifier.  A use failure requires trace
evidence that the solver read the record.  The diagnostic reference is
known useful rather than a gold retrieval label.

The downstream utility of an injected experience is conditioned on the
solver's memory-off competence rather than intrinsic to the record.
The predeclared solver contrast covers deepseek-v4-pro and
kimi-k2.7-code with all four seeds completed in both conditions.  On
\texttt{pybamm-4267}, \texttt{cfn-lint-3972}, and \texttt{aiohttp-8720}
the same supplied record moves the two solvers in opposite directions
past the two-seed margin.  In every contrast the solver with the weaker
memory-off result gains, while the stronger loses.  A retrieval
score on the record alone cannot express
this dependence, and Appendix~\ref{app:cases} traces each contrast.

Solver headroom, not a record property, predicts the direction across
the evaluation.
Table~\ref{tab:opposite-sign} profiles the 444 target and system pairs
in which the same injected record reaches both solvers.  Among the 118
pairs where both solvers move, 64 move in opposite directions.  In 61 of
those the solver with the lower memory-off count gains, and the
opposite-sign share rises with the memory-off gap.  Eight record
attributes, among them length, anchor level, and pollution rate, stay
statistically indistinguishable between opposite-sign and same-sign pairs
at a smallest permutation $p$ of 0.25, and both solver pairings that
involve glm-5 repeat the pattern.  A rule reading only memory-off
headroom predicts the sign of the paired effect on 56.5 percent of the 641
pairings changing Resolved, above the 52.4 percent of a rule reading
only record content.

\begin{table}[t]
\centering
\small
\begin{tabular*}{\linewidth}{@{\extracolsep{\fill}}lrrrr@{}}
\toprule
Memory-off gap & Pairs & Both nonzero & Opposite & Share \\
\midrule
0 seeds & 184 & 15 & 0 & 0.0\% \\
1 seed & 172 & 65 & 34 & 52.3\% \\
2 seeds & 52 & 24 & 18 & 75.0\% \\
3+ seeds & 36 & 14 & 12 & 85.7\% \\
\midrule
All & 444 & 118 & 64 & 54.2\% \\
\bottomrule
\end{tabular*}
\caption{Opposite-sign share by memory-off gap for the deepseek-v4-pro
and kimi-k2.7-code pairing.  Pairs counts the 444 target and system
pairs in each gap bucket, Both nonzero counts pairs where both solvers
move against their memory-off runs, Opposite counts pairs
moving in opposite directions, and Share divides Opposite by Both
nonzero.}
\label{tab:opposite-sign}
\end{table}

Record content separates from record volume.  The 334-line A-MEM record
of Figure~\ref{fig:record-contrast} drops \texttt{aiohttp-7907} from
four seeds to zero, while a one-line Mem0 record leaves the
same target at four of four.  Whether the content stays anchored to the
target predicts the outcome, not its length, and losses
concentrate on targets a solver resolves without memory,
consistent with the ceiling behind
Table~\ref{tab:retrieval-results}.

\subsection{Failure Attribution}

The diagnostic cohort covers the deepseek-v4-pro and kimi-k2.7-code
rosters of Table~\ref{tab:retrieval-results} and yields 231 failing
pairings over 85 targets, attributed in Table~\ref{tab:failure-attribution}.
A deterministic script assigns every label from logged artifacts.
Under the written anchor rule, form degradation
dominates while ranking misses are nearly absent, so the
systems usually reach relevant history and lose the pairing in how they
render it.  Injected records average 65 lines against a four-field frozen
record, and half trigger the lexical instruction pollution markers.

A blind LLM validity audit with two judges from model families
uninvolved in construction or solving re-labeled all 231 pairings under
the same written manual.  The audit supports the coarse attribution and
bounds the fine one.  Coverage labels match the script exactly, and the
combined share of the two stages located in the supplied record stays
dominant for every rater at 61.9 to 81.4 percent.  The boundary between
ranking and form moves with how strictly a rater reads the anchor rule,
at Cohen's kappa of 0.32 and 0.24 against the script and 0.35 between the
judges.  The form degradation share in
Table~\ref{tab:failure-attribution} therefore follows the written anchor
rule rather than rater consensus.  Appendix~\ref{app:audit} reports the
full audit.  Under every rater, the gap between availability and use
sits in the record that existing memory systems supply, not in coverage
or downstream use.

\begin{table}[t]
\centering
{\footnotesize
\setlength{\tabcolsep}{3pt}
\begin{tabularx}{\linewidth}{@{}lYrr@{}}
\toprule
First broken stage & Definition & Pairings & Share \\
\midrule
Coverage miss & source history never entered the system store & 37 & 16.0\% \\
Ranking miss & stored, but the injected record shares no anchor with it & 3 & 1.3\% \\
Form degradation & relevant history reached, but the fix pattern dropped or buried in raw transcript & 160 & 69.3\% \\
Use failure & record matches the reference on anchor and repair principle, yet the run fails & 31 & 13.4\% \\
\midrule
All failing pairings & & 231 & 100\% \\
\bottomrule
\end{tabularx}
}
\caption{Failure attribution for the retrieval condition.  A failing pairing
is a target, solver, and system whose four-seed resolved count falls below
the matched memory-off count.  Each pairing receives the first broken stage.}
\label{tab:failure-attribution}
\end{table}

\begin{table}[t]
\centering
{\small
\begin{tabular*}{\linewidth}{@{\extracolsep{\fill}}lrr@{}}
\toprule
Form degradation mechanism & Pairings & Share \\
\midrule
Instruction pollution        & 64 & 40.0\% \\
Overgeneralized fix pattern  & 29 & 18.1\% \\
Tool or shell noise          & 16 & 10.0\% \\
Schema omits fix field       & 14 &  8.8\% \\
Fix replaced by symptom      &  6 &  3.8\% \\
Other                        & 31 & 19.4\% \\
\midrule
All form degradation         & 160 & 100\% \\
\bottomrule
\end{tabular*}
}
\caption{Form degradation decomposition.  Each of the 160 form degradation
pairings in the diagnostic cohort receives its primary mechanism.  Other
groups records with metadata only, retrieval of a wrong span, truncation,
and unclear cases.}
\label{tab:stage-c}
\end{table}

\begin{figure}[t]
\centering
\includegraphics[width=\linewidth]{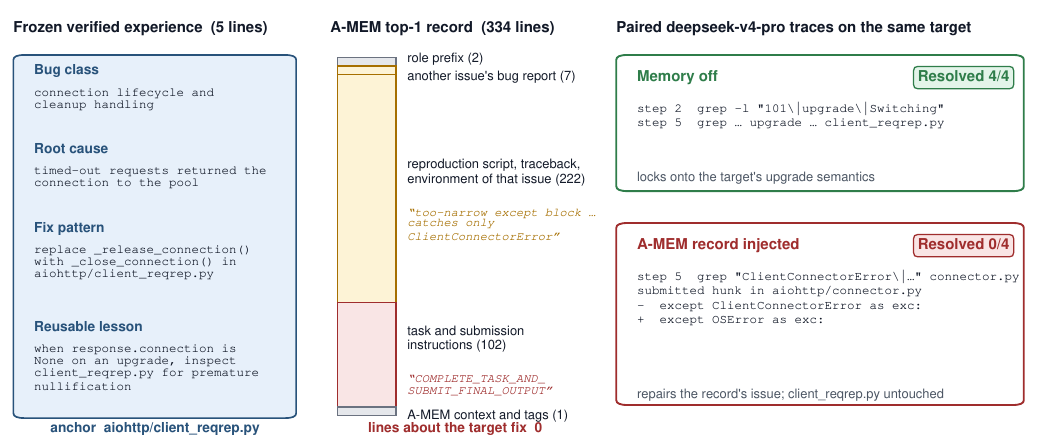}
\caption{Record form contrast on \texttt{aiohttp-7907}.  The
frozen verified experience states the fix pattern and the
\texttt{client\_reqrep.py} anchor in five lines, while the A-MEM
record spends 334 lines on another issue's report, reproduction
steps, and instructions.  In paired traces the memory-off
run locks onto the upgrade semantics and resolves every seed,
while the injected record redirects deepseek-v4-pro to the record's
except block in \texttt{connector.py} and resolves none.}
\label{fig:record-contrast}
\end{figure}

Table~\ref{tab:stage-c} decomposes form degradation by primary mechanism.
Instruction pollution and overgeneralized fix patterns together account
for over half of the cases.  Transcript systems, MemoryOS and A-MEM,
inject raw conversation chunks and produce every case of instruction
pollution and shell noise.  Atomic systems, Mem0 and SimpleMem, store
facts an LLM extracts and account for all cases of overgeneralization,
schema omission, and symptom substitution.  The break is a design choice
rather than a tuning failure.  A code audit finds that instruction
pollution crosses a task instruction boundary in only three of sixty-four
cases, so the transcript harm is verbatim storage without an extraction
filter rather than a chunk boundary, and a filter prompt would not reach
the code path that concatenates the original text.  In the atomic systems
an extraction prompt written for dialogue facts and a schema without a
repair slot collapse the fix into a generic sentence.
Figure~\ref{fig:record-contrast} shows the dominant break on one target,
and Appendix~\ref{app:mechanisms} gives the mechanism attribution.

A strip ablation tests whether instruction content causes the loss.
On sixty-two of the sixty-four instruction pollution pairings we reinject
each record
without the flagged lines, and compare against deleting equally many
unflagged lines at random.  Both arms recover seeds over the
untouched record, by 0.60 and 0.48 per pairing, and a target-level
permutation test rejects the null for each.  Because random deletion
recovers comparably, the harm operates through
transcript volume rather than instruction semantics, and the lexical
pollution flag marks a bulky record rather than a failure
mechanism.

A counterfactual substitution bounds what form repair could recover.
Replacing each failing record with the frozen verified experience for
the same target recovers 196 seeds over the 160
form degradation pairings and 78 over the 64 instruction pollution
pairings.  The two diagnosed solvers otherwise lose 93 seeds.  This bound
marks what one verified record recovers, not a prediction for a
repaired system, because frozen injection changes more than record form.

\section{Discussion}

The benchmark separates two questions, whether the frozen verified
experience helps a solver once it reaches the context, and whether
existing memory systems turn the same history into useful context.  The
frozen record witnesses what history makes available rather than
bounding it from above, so the availability-to-use gap is a
floor.  Where headroom exists, the
gap lies in the size of the gains rather than in harm.

Failure attribution and case evidence support three design implications.
All three rest on the solvers that run both evaluation layers, not on the
two transfer-only solvers.
Injection should be gated on expected solver headroom rather than applied
uniformly.  When a solver fails a target without memory and the record
carries at least a directory or identifier anchor, the record raises
Resolved on 55.2 percent of such pairings with no observed loss, while at
a four of four ceiling even the frozen verified experience lowers Resolved
on 25.7 percent of pairings.  The dominant break is the form of the
transcript record, and the ablation places its harm in raw volume, so a
system should compress a retrieved trajectory to the span stating the
fix rather than filter instruction text by surface markers.  Preserving a
fix field is necessary but not sufficient, because records keeping the
fix pattern do not lower the loss rate, so an extraction schema must
retain the constraints anchoring the fix to the target.

Selection creates one boundary.  The offline patch-pattern screen compares
target and history gold patches, and the 111
targets have positive deepseek-v4-flash reference uplift by design.  This
suits a benchmark of the downstream value of known transferable
experience, not an estimate of how often a deployed agent finds
such a pair.
The retrieval-cohort ceiling reflects solver strength rather than a lost
opportunity.  All 140 ceiling target and solver pairs of that cohort stay
below four of four under the reference solver, so the reference setting
retained them under the uplift rule and the ceiling appears only after
transfer.

\section{Limitations and Conclusion}

VibeMemBench measures persistent memory under a declared agent
configuration and accounting rule, over 111 targets from an
oracle-assisted patch-pattern screen and a positive reference uplift
rule that supply no gold retrieval label.
The solver pool omits proprietary frontier models such as the Claude and
GPT families for reasons of inference budget, so the solver-agnostic
protocol stays untested for them.
The release supplies the task histories, executable predicate, and
intervention protocol with full provenance.

\bibliography{references}

@inproceedings{yang2026talk2code,
  author       = {Weibin Yang and Liangru Xie and Jieyun Cai and Yuxiang Yan and Hong{-}Ning Dai and Hao Wang},
  editor       = {Sven Koenig and Chad Jenkins and Matthew E. Taylor},
  title        = {{Talk2Code:} {A} Multi-Turn Interaction Benchmark with Dual-Track Evaluation for Code Generation},
  booktitle    = {Fortieth {AAAI} Conference on Artificial Intelligence, Thirty-Eighth Conference on Innovative Applications of Artificial Intelligence, Sixteenth Symposium on Educational Advances in Artificial Intelligence, {AAAI} 2026, Singapore, January 20-27, 2026},
  pages        = {34331--34339},
  publisher    = {{AAAI} Press},
  year         = {2026},
  url          = {https://doi.org/10.1609/aaai.v40i40.40730},
  doi          = {10.1609/AAAI.V40I40.40730},
  bibsource    = {dblp computer science bibliography, https://dblp.org}
}

@inproceedings{ni2026gittaskbench,
  author       = {Ziyi Ni and Huacan Wang and Shuo Zhang and Shuo Lu and Ziyang He and Wang You and Zhenheng Tang and Sen Hu and Bo Li and Chen Hu and Binxing Jiao and Daxin Jiang and Yuntao Du and Pin Lyu},
  editor       = {Sven Koenig and Chad Jenkins and Matthew E. Taylor},
  title        = {{GitTaskBench:} {A} Benchmark for Code Agents Solving Real-World Tasks Through Code Repository Leveraging},
  booktitle    = {Fortieth {AAAI} Conference on Artificial Intelligence, Thirty-Eighth Conference on Innovative Applications of Artificial Intelligence, Sixteenth Symposium on Educational Advances in Artificial Intelligence, {AAAI} 2026, Singapore, January 20-27, 2026},
  pages        = {32564--32572},
  publisher    = {{AAAI} Press},
  year         = {2026},
  url          = {https://doi.org/10.1609/aaai.v40i38.40533},
  doi          = {10.1609/AAAI.V40I38.40533},
  bibsource    = {dblp computer science bibliography, https://dblp.org}
}

@inproceedings{ouyang2026dscodebench,
  author       = {Shuyin Ouyang and Dong Huang and Jingwen Guo and Zeyu Sun and Qihao Zhu and Jie M. Zhang},
  editor       = {Sven Koenig and Chad Jenkins and Matthew E. Taylor},
  title        = {{DSCodeBench:} {A} Realistic Benchmark for Data Science Code Generation},
  booktitle    = {Fortieth {AAAI} Conference on Artificial Intelligence, Thirty-Eighth Conference on Innovative Applications of Artificial Intelligence, Sixteenth Symposium on Educational Advances in Artificial Intelligence, {AAAI} 2026, Singapore, January 20-27, 2026},
  pages        = {32628--32636},
  publisher    = {{AAAI} Press},
  year         = {2026},
  url          = {https://doi.org/10.1609/aaai.v40i38.40540},
  doi          = {10.1609/AAAI.V40I38.40540},
  bibsource    = {dblp computer science bibliography, https://dblp.org}
}

@inproceedings{dai2026memoryart,
  author       = {Renke Dai and Hebin Hu and Jiahui Zhang and Yilin Kang and Ah{-}Hwee Tan},
  editor       = {Sven Koenig and Chad Jenkins and Matthew E. Taylor},
  title        = {{MemoryART:} Enhancing {LLM}s via Multi-Memory Models with Adaptive Resonance Theory for Healthcare Agents},
  booktitle    = {Fortieth {AAAI} Conference on Artificial Intelligence, Thirty-Eighth Conference on Innovative Applications of Artificial Intelligence, Sixteenth Symposium on Educational Advances in Artificial Intelligence, {AAAI} 2026, Singapore, January 20-27, 2026},
  pages        = {20676--20683},
  publisher    = {{AAAI} Press},
  year         = {2026},
  url          = {https://doi.org/10.1609/aaai.v40i25.39205},
  doi          = {10.1609/AAAI.V40I25.39205},
  bibsource    = {dblp computer science bibliography, https://dblp.org}
}

@inproceedings{ye2026realwebassist,
  author       = {Suyu Ye and Haojun Shi and Darren Shih and Hyokun Yun and Tanya G. Roosta and Tianmin Shu},
  editor       = {Sven Koenig and Chad Jenkins and Matthew E. Taylor},
  title        = {{RealWebAssist:} {A} Benchmark for Long-Horizon Web Assistance with Real-World Users},
  booktitle    = {Fortieth {AAAI} Conference on Artificial Intelligence, Thirty-Eighth Conference on Innovative Applications of Artificial Intelligence, Sixteenth Symposium on Educational Advances in Artificial Intelligence, {AAAI} 2026, Singapore, January 20-27, 2026},
  pages        = {34441--34449},
  publisher    = {{AAAI} Press},
  year         = {2026},
  url          = {https://doi.org/10.1609/aaai.v40i40.40742},
  doi          = {10.1609/AAAI.V40I40.40742},
  bibsource    = {dblp computer science bibliography, https://dblp.org}
}

@inproceedings{yang2026probench,
  author       = {Leyang Yang and Ziwei Wang and Xiaoxuan Tang and Sheng Zhou and Dajun Chen and Wei Jiang and Yong Li},
  editor       = {Sven Koenig and Chad Jenkins and Matthew E. Taylor},
  title        = {{ProBench:} Benchmarking {GUI} Agents with Accurate Process Information},
  booktitle    = {Fortieth {AAAI} Conference on Artificial Intelligence, Thirty-Eighth Conference on Innovative Applications of Artificial Intelligence, Sixteenth Symposium on Educational Advances in Artificial Intelligence, {AAAI} 2026, Singapore, January 20-27, 2026},
  pages        = {27547--27555},
  publisher    = {{AAAI} Press},
  year         = {2026},
  url          = {https://doi.org/10.1609/aaai.v40i32.39974},
  doi          = {10.1609/AAAI.V40I32.39974},
  bibsource    = {dblp computer science bibliography, https://dblp.org}
}

@inproceedings{yuan2025dmtrolebench,
  author       = {Dingbo Yuan and Yipeng Chen and Guodong Liu and Chenchen Li and Chengfu Tang and Dongxu Zhang and Zhenkui Wang and Xudong Wang and Song Liu},
  editor       = {Toby Walsh and Julie Shah and Zico Kolter},
  title        = {{DMT-RoleBench:} {A} Dynamic Multi-Turn Dialogue Based Benchmark for Role-Playing Evaluation of Large Language Model and Agent},
  booktitle    = {Thirty-Ninth {AAAI} Conference on Artificial Intelligence, Thirty-Seventh Conference on Innovative Applications of Artificial Intelligence, Fifteenth Symposium on Educational Advances in Artificial Intelligence, {AAAI} 2025, Philadelphia, PA, USA, February 25 - March 4, 2025},
  pages        = {25760--25768},
  publisher    = {{AAAI} Press},
  year         = {2025},
  url          = {https://doi.org/10.1609/aaai.v39i24.34768},
  doi          = {10.1609/AAAI.V39I24.34768},
  bibsource    = {dblp computer science bibliography, https://dblp.org}
}

@inproceedings{zhang2025hierarchicalcontextpruning,
  author       = {Lei Zhang and Yunshui Li and Jiaming Li and Xiaobo Xia and Jiaxi Yang and Run Luo and Minzheng Wang and Longze Chen and Junhao Liu and Qiang Qu and Min Yang},
  editor       = {Toby Walsh and Julie Shah and Zico Kolter},
  title        = {Hierarchical Context Pruning: Optimizing Real-World Code Completion with Repository-Level Pretrained Code {LLM}s},
  booktitle    = {Thirty-Ninth {AAAI} Conference on Artificial Intelligence, Thirty-Seventh Conference on Innovative Applications of Artificial Intelligence, Fifteenth Symposium on Educational Advances in Artificial Intelligence, {AAAI} 2025, Philadelphia, PA, USA, February 25 - March 4, 2025},
  pages        = {25886--25894},
  publisher    = {{AAAI} Press},
  year         = {2025},
  url          = {https://doi.org/10.1609/aaai.v39i24.34782},
  doi          = {10.1609/AAAI.V39I24.34782},
  bibsource    = {dblp computer science bibliography, https://dblp.org}
}

@inproceedings{zhong2024memorybank,
  author       = {Wanjun Zhong and Lianghong Guo and Qiqi Gao and He Ye and Yanlin Wang},
  editor       = {Michael J. Wooldridge and Jennifer G. Dy and Sriraam Natarajan},
  title        = {{MemoryBank:} Enhancing Large Language Models with Long-Term Memory},
  booktitle    = {Thirty-Eighth {AAAI} Conference on Artificial Intelligence, {AAAI} 2024, Thirty-Sixth Conference on Innovative Applications of Artificial Intelligence, {IAAI} 2024, Fourteenth Symposium on Educational Advances in Artificial Intelligence, {EAAI} 2024, February 20-27, 2024, Vancouver, Canada},
  pages        = {19724--19731},
  publisher    = {{AAAI} Press},
  year         = {2024},
  url          = {https://doi.org/10.1609/aaai.v38i17.29946},
  doi          = {10.1609/AAAI.V38I17.29946},
  bibsource    = {dblp computer science bibliography, https://dblp.org}
}

@inproceedings{badertdinov2026swerebenchv2languageagnosticswe,
  author       = {Ibragim Badertdinov and Maksim Nekrashevich and Anton Shevtsov and Alexander Golubev},
  title        = {{SWE-rebench V2}: Language-Agnostic {SWE} Task Collection at Scale},
  booktitle    = {Forty-third International Conference on Machine Learning, {ICML} 2026, Seoul, South Korea, July 6-11, 2026},
  year         = {2026},
  url          = {https://openreview.net/forum?id=UCAda9kS57},
  doi          = {10.48550/arXiv.2602.23866},
  eprinttype   = {arXiv},
  eprint       = {2602.23866},
  note         = {ICML 2026; arXiv:2602.23866. DBLP record not available at time of bibliography audit}
}

@misc{zhang2025qwen3embedding,
  author       = {Yanzhao Zhang and
                  Mingxin Li and
                  Dingkun Long and
                  Xin Zhang and
                  Huan Lin and
                  Baosong Yang and
                  Pengjun Xie and
                  An Yang and
                  Dayiheng Liu and
                  Junyang Lin and
                  Fei Huang and
                  Jingren Zhou},
  title        = {{Qwen3 Embedding:} Advancing Text Embedding and Reranking Through Foundation Models},
  year         = {2025},
  url          = {https://doi.org/10.48550/arXiv.2506.05176},
  doi          = {10.48550/ARXIV.2506.05176},
  archivePrefix = {arXiv},
  eprint       = {2506.05176},
  bibsource    = {dblp computer science bibliography, https://dblp.org}
}

@misc{deepseek2026models,
  author       = {{DeepSeek AI}},
  title        = {{DeepSeek API} Models and Pricing Documentation},
  year         = {2026},
  howpublished = {DeepSeek API Documentation},
  url          = {https://api-docs.deepseek.com/quick_start/pricing/},
  note         = {Model IDs \texttt{deepseek-v4-flash} and \texttt{deepseek-v4-pro}. Accessed 2026-07-20}
}

@misc{glm2026glm5,
  author       = {GLM},
  title        = {{GLM-5:} from Vibe Coding to Agentic Engineering},
  year         = {2026},
  url          = {https://doi.org/10.48550/arXiv.2602.15763},
  doi          = {10.48550/ARXIV.2602.15763},
  archivePrefix = {arXiv},
  eprint       = {2602.15763},
  bibsource    = {dblp computer science bibliography, https://dblp.org}
}

@misc{zai2026glm52docs,
  author       = {{Z.AI}},
  title        = {{GLM-5.2} Developer Documentation},
  year         = {2026},
  howpublished = {Z.AI Developer Documentation},
  url          = {https://docs.z.ai/guides/llm/glm-5.2},
  note         = {Model ID \texttt{glm-5.2}. Accessed 2026-07-20}
}

@misc{kimi2025k2,
  author       = {{Kimi Team}},
  title        = {Kimi {K2:} Open Agentic Intelligence},
  year         = {2025},
  url          = {https://doi.org/10.48550/arXiv.2507.20534},
  doi          = {10.48550/ARXIV.2507.20534},
  archivePrefix = {arXiv},
  eprint       = {2507.20534},
  bibsource    = {dblp computer science bibliography, https://dblp.org}
}

@misc{moonshot2026kimi27codedocs,
  author       = {{Moonshot AI}},
  title        = {{Kimi K2.7 Code} Quickstart Documentation},
  year         = {2026},
  howpublished = {Kimi API Platform Documentation},
  url          = {https://platform.kimi.ai/docs/guide/kimi-k2-7-code-quickstart},
  note         = {Model ID \texttt{kimi-k2.7-code}. Accessed 2026-07-20}
}

@misc{qwen2026qwen38max,
  author       = {{Qwen Team}},
  title        = {{Qwen3.8-Max:} A New Bar for Coding and Cowork},
  year         = {2026},
  howpublished = {Qwen release announcement},
  url          = {https://qwen.ai/blog?id=qwen3.8},
  note         = {Released 2026-08-03. Evaluated model identifier \texttt{qwen3.8-max}. Accessed 2026-08-13}
}

@inproceedings{jimenez2024swebench,
  author       = {Carlos E. Jimenez and John Yang and Alexander Wettig and Shunyu Yao and Kexin Pei and Ofir Press and Karthik R. Narasimhan},
  title        = {{SWE}-bench: Can Language Models Resolve Real-world {G}ithub Issues?},
  booktitle    = {The Twelfth International Conference on Learning Representations, {ICLR} 2024, Vienna, Austria, May 7-11, 2024},
  publisher    = {OpenReview.net},
  year         = {2024},
  url          = {https://openreview.net/forum?id=VTF8yNQM66},
  bibsource    = {dblp computer science bibliography, https://dblp.org}
}

@inproceedings{maharana2024locomo,
  author       = {Adyasha Maharana and Dong{-}Ho Lee and Sergey Tulyakov and Mohit Bansal and Francesco Barbieri and Yuwei Fang},
  editor       = {Lun{-}Wei Ku and Andre Martins and Vivek Srikumar},
  title        = {Evaluating Very Long-Term Conversational Memory of {LLM} Agents},
  booktitle    = {Proceedings of the 62nd Annual Meeting of the Association for Computational Linguistics (Volume 1: Long Papers), {ACL} 2024, Bangkok, Thailand, August 11-16, 2024},
  pages        = {13851--13870},
  publisher    = {Association for Computational Linguistics},
  year         = {2024},
  url          = {https://doi.org/10.18653/v1/2024.acl-long.747},
  doi          = {10.18653/V1/2024.ACL-LONG.747},
  bibsource    = {dblp computer science bibliography, https://dblp.org}
}

@inproceedings{wu2025longmemeval,
  author       = {Di Wu and Hongwei Wang and Wenhao Yu and Yuwei Zhang and Kai{-}Wei Chang and Dong Yu},
  title        = {{LongMemEval:} Benchmarking Chat Assistants on Long-Term Interactive Memory},
  booktitle    = {The Thirteenth International Conference on Learning Representations, {ICLR} 2025, Singapore, April 24-28, 2025},
  publisher    = {OpenReview.net},
  year         = {2025},
  url          = {https://openreview.net/forum?id=pZiyCaVuti},
  bibsource    = {dblp computer science bibliography, https://dblp.org}
}

@inproceedings{chhikara2025mem0,
  author       = {Prateek Chhikara and Dev Khant and Saket Aryan and Taranjeet Singh and Deshraj Yadav},
  editor       = {In{\^{e}}s Lynce and Nello Murano and Mauro Vallati and Serena Villata and Federico Chesani and Michela Milano and Andrea Omicini and Mehdi Dastani},
  title        = {{Mem0:} Building Production-Ready {AI} Agents with Scalable Long-Term Memory},
  booktitle    = {{ECAI} 2025 - 28th European Conference on Artificial Intelligence, 25-30 October 2025, Bologna, Italy - Including 14th Conference on Prestigious Applications of Intelligent Systems {(PAIS} 2025)},
  series       = {Frontiers in Artificial Intelligence and Applications},
  volume       = {413},
  pages        = {2993--3000},
  publisher    = {{IOS} Press},
  year         = {2025},
  url          = {https://doi.org/10.3233/FAIA251160},
  doi          = {10.3233/FAIA251160},
  bibsource    = {dblp computer science bibliography, https://dblp.org}
}

@misc{xu2025amem,
  author       = {Wujiang Xu and Zujie Liang and Kai Mei and Hang Gao and Juntao Tan and Yongfeng Zhang},
  title        = {{A-MEM:} Agentic Memory for {LLM} Agents},
  year         = {2025},
  url          = {https://doi.org/10.48550/arXiv.2502.12110},
  doi          = {10.48550/ARXIV.2502.12110},
  archivePrefix = {arXiv},
  eprint       = {2502.12110},
  bibsource    = {dblp computer science bibliography, https://dblp.org}
}

@misc{fang2025lightmem,
  author       = {Jizhan Fang and Xinle Deng and Haoming Xu and Ziyan Jiang and Yuqi Tang and Ziwen Xu and Shumin Deng and Yunzhi Yao and Mengru Wang and Shuofei Qiao and Huajun Chen and Ningyu Zhang},
  title        = {{LightMem:} Lightweight and Efficient Memory-Augmented Generation},
  year         = {2025},
  url          = {https://doi.org/10.48550/arXiv.2510.18866},
  doi          = {10.48550/ARXIV.2510.18866},
  archivePrefix = {arXiv},
  eprint       = {2510.18866},
  bibsource    = {dblp computer science bibliography, https://dblp.org}
}

@inproceedings{kang2025memoryos,
  author       = {Jiazheng Kang and Mingming Ji and Zhe Zhao and Ting Bai},
  editor       = {Christos Christodoulopoulos and Tanmoy Chakraborty and Carolyn Rose and Violet Peng},
  title        = {Memory {OS} of {AI} Agent},
  booktitle    = {Proceedings of the 2025 Conference on Empirical Methods in Natural Language Processing, {EMNLP} 2025, Suzhou, China, November 4-9, 2025},
  pages        = {25961--25970},
  publisher    = {Association for Computational Linguistics},
  year         = {2025},
  url          = {https://doi.org/10.18653/v1/2025.emnlp-main.1318},
  doi          = {10.18653/V1/2025.EMNLP-MAIN.1318},
  bibsource    = {dblp computer science bibliography, https://dblp.org}
}

@misc{liu2026simplemem,
  author       = {Jiaqi Liu and Yaofeng Su and Peng Xia and Siwei Han and Zeyu Zheng and Cihang Xie and Mingyu Ding and Huaxiu Yao},
  title        = {{SimpleMem:} Efficient Lifelong Memory for {LLM} Agents},
  year         = {2026},
  url          = {https://doi.org/10.48550/arXiv.2601.02553},
  doi          = {10.48550/ARXIV.2601.02553},
  archivePrefix = {arXiv},
  eprint       = {2601.02553},
  bibsource    = {dblp computer science bibliography, https://dblp.org}
}

@misc{joshi2025swebenchcl,
  author       = {Thomas Joshi and Shayan Chowdhury and Fatih Uysal},
  title        = {{SWE-Bench-CL:} Continual Learning for Coding Agents},
  year         = {2025},
  url          = {https://doi.org/10.48550/arXiv.2507.00014},
  doi          = {10.48550/ARXIV.2507.00014},
  archivePrefix = {arXiv},
  eprint       = {2507.00014},
  bibsource    = {dblp computer science bibliography, https://dblp.org}
}

@inproceedings{tan2025membench,
  author       = {Haoran Tan and
                  Zeyu Zhang and
                  Chen Ma and
                  Xu Chen and
                  Quanyu Dai and
                  Zhenhua Dong},
  editor       = {Wanxiang Che and
                  Joyce Nabende and
                  Ekaterina Shutova and
                  Mohammad Taher Pilehvar},
  title        = {{MemBench:} Towards More Comprehensive Evaluation on the Memory of {LLM}-based
                  Agents},
  booktitle    = {Findings of the Association for Computational Linguistics, {ACL} 2025,
                  Vienna, Austria, July 27 - August 1, 2025},
  series       = {Findings of {ACL}},
  volume       = {{ACL} 2025},
  pages        = {19336--19352},
  publisher    = {Association for Computational Linguistics},
  year         = {2025},
  url          = {https://doi.org/10.18653/v1/2025.findings-acl.989},
  doi          = {10.18653/V1/2025.FINDINGS-ACL.989},
  bibsource    = {dblp computer science bibliography, https://dblp.org}
}

@inproceedings{liu2024repobench,
  author       = {Tianyang Liu and
                  Canwen Xu and
                  Julian J. McAuley},
  title        = {{RepoBench:} Benchmarking Repository-Level Code Auto-Completion Systems},
  booktitle    = {The Twelfth International Conference on Learning Representations,
                  {ICLR} 2024, Vienna, Austria, May 7-11, 2024},
  publisher    = {OpenReview.net},
  year         = {2024},
  url          = {https://openreview.net/forum?id=pPjZIOuQuF},
  bibsource    = {dblp computer science bibliography, https://dblp.org}
}

@inproceedings{wang2024memoryllm,
  author       = {Yu Wang and
                  Yifan Gao and
                  Xiusi Chen and
                  Haoming Jiang and
                  Shiyang Li and
                  Jingfeng Yang and
                  Qingyu Yin and
                  Zheng Li and
                  Xian Li and
                  Bing Yin and
                  Jingbo Shang and
                  Julian J. McAuley},
  editor       = {Ruslan Salakhutdinov and
                  Zico Kolter and
                  Katherine A. Heller and
                  Adrian Weller and
                  Nuria Oliver and
                  Jonathan Scarlett and
                  Felix Berkenkamp},
  title        = {{MEMORYLLM:} Towards Self-Updatable Large Language Models},
  booktitle    = {Forty-first International Conference on Machine Learning, {ICML} 2024,
                  Vienna, Austria, July 21-27, 2024},
  series       = {Proceedings of Machine Learning Research},
  volume       = {235},
  pages        = {50453--50466},
  publisher    = {{PMLR} / OpenReview.net},
  year         = {2024},
  url          = {https://proceedings.mlr.press/v235/wang24s.html},
  bibsource    = {dblp computer science bibliography, https://dblp.org}
}

@misc{packer2023memgpt,
  author       = {Charles Packer and
                  Vivian Fang and
                  Shishir G. Patil and
                  Kevin Lin and
                  Sarah Wooders and
                  Joseph E. Gonzalez},
  title        = {{MemGPT:} Towards {LLM}s as Operating Systems},
  year         = {2023},
  url          = {https://doi.org/10.48550/arXiv.2310.08560},
  doi          = {10.48550/ARXIV.2310.08560},
  archivePrefix = {arXiv},
  eprint       = {2310.08560},
  bibsource    = {dblp computer science bibliography, https://dblp.org}
}

@inproceedings{zhang2023repocoder,
  author       = {Fengji Zhang and
                  Bei Chen and
                  Yue Zhang and
                  Jacky Keung and
                  Jin Liu and
                  Daoguang Zan and
                  Yi Mao and
                  Jian{-}Guang Lou and
                  Weizhu Chen},
  editor       = {Houda Bouamor and
                  Juan Pino and
                  Kalika Bali},
  title        = {{RepoCoder:} Repository-Level Code Completion Through Iterative Retrieval
                  and Generation},
  booktitle    = {Proceedings of the 2023 Conference on Empirical Methods in Natural
                  Language Processing, {EMNLP} 2023, Singapore, December 6-10, 2023},
  pages        = {2471--2484},
  publisher    = {Association for Computational Linguistics},
  year         = {2023},
  url          = {https://doi.org/10.18653/v1/2023.emnlp-main.151},
  doi          = {10.18653/V1/2023.EMNLP-MAIN.151},
  bibsource    = {dblp computer science bibliography, https://dblp.org}
}

@inproceedings{yang2024sweagent,
  author       = {John Yang and
                  Carlos E. Jimenez and
                  Alexander Wettig and
                  Kilian Lieret and
                  Shunyu Yao and
                  Karthik Narasimhan and
                  Ofir Press},
  editor       = {Amir Globersons and
                  Lester Mackey and
                  Danielle Belgrave and
                  Angela Fan and
                  Ulrich Paquet and
                  Jakub M. Tomczak and
                  Cheng Zhang},
  title        = {{SWE}-agent: Agent-Computer Interfaces Enable Automated Software Engineering},
  booktitle    = {Advances in Neural Information Processing Systems 37: Annual Conference
                  on Neural Information Processing Systems 2024, NeurIPS 2024, Vancouver,
                  BC, Canada, December 10 - 15, 2024},
  year         = {2024},
  url          = {http://papers.nips.cc/paper\_files/paper/2024/hash/5a7c947568c1b1328ccc5230172e1e7c-Abstract-Conference.html},
  bibsource    = {dblp computer science bibliography, https://dblp.org}
}

\clearpage
\appendix
\section*{Technical Appendix}
\suppressfloats[t]

This appendix holds the source-field dictionary, the construction and
condition contracts, the full exclusion and audit material, and the
extended breakdowns behind the main-paper claims.  Section numbers
restart with letters, while table and figure numbers continue from the
main paper.

\section{Related-Work Contrast}
\label{app:related-work}

Table~\ref{tab:related-work} separates task success from memory probes on
five auditable axes and records whether prior work isolates a memory
intervention under matched conditions.  Only the last row combines an
executable task outcome with a matched memory toggle, and no family
above it supplies both.

\begin{table}[t]
\centering
\small
\begin{tabularx}{\linewidth}{@{}L{0.16\linewidth}YYL{0.16\linewidth}L{0.15\linewidth}@{}}
\toprule
Family & Task unit & Primary signal & Memory target supplied & Matched memory toggle \\
\midrule
Coding benchmarks & Repository issue or completion & Tests or task-specific check & No & No \\
Memory benchmarks & Interaction history and query & Answer or capability score & Query-targeted evidence & No \\
Memory systems and continual coding & Update stream or ordered task sequence & System or continual-learning metric & Not a benchmark label & No \\
VibeMemBench & Prior coding history and target issue & Executable tests, solver tokens, and steps & No instance-level target & Yes \\
\bottomrule
\end{tabularx}
\caption{Related-work contrast.  Rows group prior work into four
families; columns record the task unit, primary evaluation signal,
whether a memory target is supplied, and whether a matched memory
toggle is isolated.}
\label{tab:related-work}
\end{table}

\section{Construction Audit}
\label{app:construction}

Table~\ref{tab:construction-audit} lists each stage of the SIEVE
construction, its rule, and the retained artifact.  SIEVE names the
successive filtering stages and is not an acronym.  The release manifest
records the script variant, prompt, chronology, and count fields of each
stage.

\begin{table}[t]
\centering
\small
\begin{tabularx}{\linewidth}{@{}L{0.22\linewidth}Y@{}}
\toprule
Stage & Rule and retained artifact \\
\midrule
Patch validation & Inject the source gold patch and execute the declared
validation.  Discard an instance when the gold patch does not pass. \\
Patch-pattern screening & Offline construction compares a target gold patch to
earlier history patches from the same repository.  Structural screening and an LLM
return evidence-backed candidate pairs. \\
History execution & Run relevant history with MiniSWEAgent,
deepseek-v4-flash,
one seed, temperature 0.5, and a 250-step limit. \\
Target-aware experience & Use the target problem statement to focus a summary
of the history trajectory and history final patch.  The target gold and test
patches are not inputs. \\
Target retention & Keep a target when memory-off is below 4/4 and the
maximum-uplift candidate memory improves the deepseek-v4-flash reference
setting. \\
Packaging & Retain the base commit, image, installation configuration, test
command, and executable result for every released target. \\
\bottomrule
\end{tabularx}
\caption{Construction audit.  Left column names each SIEVE stage;
right column gives the rule applied and the artifact retained at
that stage.}
\label{tab:construction-audit}
\end{table}

\subsection{Patch-Pattern Scoring Rule}

For a candidate target $t$, the offline screen first admits only earlier
instances from the same repository and excludes every target identifier.  It
extracts patch file paths $F$, directories $D$, and cleaned context or
definition identifiers $I$.  Let $\delta_f(t,h)$ equal one for a shared file.
Let $\delta_d(t,h)$ equal one only when there is no shared file but there is a
shared directory.  The common deterministic score is
\begin{equation}
r(t,h)=3\delta_f(t,h)+\delta_d(t,h)+\min(2,\left|I_t\cap I_h\right|).
\end{equation}
The score favors shared files, then shared symbols and shared directories.  A
score of at least one reaches the shortlist.  The implementation retains at
most 24 candidates in descending score and recency order, then asks an LLM to
judge each repair pattern with structured evidence.  The release manifest
names the semantic prompt variant, strict or relaxed, the model revision,
and the resulting match file.

\subsection{Target-Aware Experience Construction}

For each retained pair, the target-aware experience construction stage
produces one experience.  It receives the target problem statement, the
history problem statement, a compressed history trajectory, and the history
final patch.  The target statement only selects which historical
lesson to emphasize.  Each concrete claim is grounded in the historical
trajectory or patch.  The resulting record contains a bug class, root cause,
fix pattern, and reusable lesson.  The default script gates positive examples
on an \texttt{all\_ok} history evaluation and skips unresolved trajectories
unless the run manifest declares otherwise.  Target gold patches, target test
patches, and target test outcomes are excluded from this construction.

\section{Source Instance Fields}
\label{app:source-fields}

SWE-rebench V2 records an instance identifier, repository, base commit, gold
patch, test patch, problem statement, pull-request description, creation time,
Docker image, language, code interface, repository license, failing-to-passing
tests, passing-to-passing tests, installation configuration, and metadata.
These fields provide
the source material for VibeMemBench's provenance and execution contract.  The
released benchmark states which fields reach the solver.  In particular,
the target gold patch enters only the oracle-assisted offline patch-pattern
screen.  The target test patch remains evaluator-only.  Neither artifact may
reach target-aware experience construction, runtime memory retrieval, or the
target solver context.

Table~\ref{tab:source-record} turns the field list into an auditable
boundary.  It distinguishes source-task material from metadata and
evaluator-controlled artifacts.  The release manifest states any
additional solver-visible field and whether executable test identifiers
are visible, and gold patches and test patches remain hidden.  The
manifest also states the history cutoff for each target, so the history
set stays a source of observations rather than an answer key.

\begin{table}[t]
\centering
\small
\begin{tabularx}{\linewidth}{@{}L{0.20\linewidth}YL{0.30\linewidth}@{}}
\toprule
Field group & Fields & Visibility boundary \\
\midrule
Task and execution material & \texttt{instance\_id}, \texttt{repo},
\texttt{base\_commit}, \texttt{problem\_statement}, \texttt{image\_name},
\texttt{language}, \texttt{install\_config} & Release manifest declares
solver visibility. \\
Provenance and metadata & \texttt{pr\_description}, \texttt{created\_at},
\texttt{interface}, \texttt{license}, \texttt{meta} & Invisible by
default.  Any exposure requires an explicit audit. \\
Evaluator-controlled artifacts & \texttt{patch}, \texttt{test\_patch},
\texttt{FAIL\_TO\_PASS}, \texttt{PASS\_TO\_PASS} & Gold patch enters only
the offline patch-pattern screen.  Test patch stays evaluator-only. \\
\bottomrule
\end{tabularx}
\caption{Source-field roles in a VibeMemBench record.  Each row names a
field group, lists its SWE-rebench V2 fields, and states the visibility
boundary that the release enforces.}
\label{tab:source-record}
\end{table}

\section{Task-Type and Language Audit}
\label{app:task-type}

Table~\ref{tab:task-type-audit} reports the exact counts behind the
main-paper profile charts.  A construction audit assigns each of the 111
targets one primary task type from an eleven-way taxonomy using the
problem statement, construction-only patch evidence, and declared tests.
The audit artifacts never reach the solver.

\begin{table}[t]
\centering
\small
\begin{tabular*}{\linewidth}{@{\extracolsep{\fill}}lrr@{}}
\toprule
Primary task type & Targets & Share \\
\midrule
Bug fix & 33 & 29.7\% \\
Feature addition & 23 & 20.7\% \\
Parser or serialization logic & 16 & 14.4\% \\
Error handling or validation & 9 & 8.1\% \\
Compatibility fix & 8 & 7.2\% \\
Behavior change & 8 & 7.2\% \\
API or interface adjustment & 6 & 5.4\% \\
Configuration or dependency update & 3 & 2.7\% \\
Refactor or cleanup & 2 & 1.8\% \\
Test infrastructure or CI & 2 & 1.8\% \\
Documentation-related code change & 1 & 0.9\% \\
\midrule
All targets & 111 & 100\% \\
\bottomrule
\end{tabular*}
\caption{Primary task-type audit of the 111 targets.  Each target
receives one mutually exclusive primary label.}
\label{tab:task-type-audit}
\end{table}

The language distribution counts Python 56, JavaScript 16, Go 12, Rust
10, TypeScript 5, Kotlin 4, Swift 4, Java 2, and PHP 2.  The main-paper
figure groups the five smallest languages as Other.

\section{Reference-Setting Selection Audit}
\label{app:selection}

Table~\ref{tab:selection-audit} reports the deepseek-v4-flash reference
comparison.  Injecting the selected experience raises passing runs from
192 to 349 of 444 while lowering solver tokens and steps.  This
comparison participates in target selection.  It is a selection audit, not
evidence that the same uplift holds on an unselected source population or on
another solver.  Cross-model transfer in the main paper uses the frozen
verified experience and does not reselect a memory combination for any evaluation
model.

\begin{table}[t]
\centering
\small
\begin{tabular*}{\linewidth}{@{\extracolsep{\fill}}lrrrr@{}}
\toprule
Condition & Pass & Resolved & In/out & Steps \\
\midrule
Memory off & 192 & 43.2\% & 882M/9.4M & 24,618 \\
Frozen verified exp. & 349 & 78.6\% & 714M/8.6M & 22,200 \\
\bottomrule
\end{tabular*}
\caption{Reference-setting selection audit on the 111 outcome-selected
targets.  Pass counts passing runs out of 444 target-seed runs; In and
out are solver-side token totals; Steps counts declared agent steps
over the 444 runs.}
\label{tab:selection-audit}
\end{table}

\section{Condition Contract}
\label{app:contract}

Table~\ref{tab:condition-contract} states the invariants that both memory
conditions share and the fields that the experiment manifest fills for
each reported run.  Without these invariants, a performance difference could reflect a
changed prompt, tool, or context window rather than memory.  The
bracketed cells are the only places where the two conditions may
differ, and each names a manifest field rather than a free choice.

\begin{table}[t]
\centering
\small
\begin{tabularx}{\linewidth}{@{}L{0.24\linewidth}YY@{}}
\toprule
Field & Memory off & Memory on \\
\midrule
Model, agent, tools, sandbox & Same declared configuration & Same declared configuration \\
Task order and resource limits & Same target and seed order & Same target and seed order \\
Context budget & [fixed budget] & [the same fixed budget] \\
Persistent writes & Disabled & [write policy] \\
Retrieval and insertion & No persistent records & [trigger, ranker, truncation] \\
Resource logging & Solver tokens and steps & Solver tokens and declared steps \\
\bottomrule
\end{tabularx}
\caption{Condition contract.  Each row names a controlled field;
the second and third columns give its value in the memory-off and
memory-on conditions.  Square brackets mark fields the experiment
manifest fills for each reported run.}
\label{tab:condition-contract}
\end{table}

\section{Predeclared Ablation Schema}
\label{app:ablation-schema}

Table~\ref{tab:ablation-schema} predeclares the ablation questions.  A row
enters the reported results only when the named component exists in the
evaluated implementation and all fixed fields are logged.

\begin{table}[t]
\centering
\footnotesize
\begin{tabularx}{\linewidth}{@{}L{0.24\linewidth}L{0.30\linewidth}Y@{}}
\toprule
Question & Changed component & Fixed comparison contract \\
\midrule
Persistent memory effect & Memory off or declared memory on & $A$, $\mathcal{P}$, context budget, task order, executable predicate \\
Context-length control & Selected history or token-matched irrelevant text & $A$, token allocation, target and seed pair, insertion position \\
System mechanism & One implemented write, retrieval, or reranking component & $A$, $\mathcal{P}$, remaining memory pipeline, solver-token and step logging \\
Solver use & Memory condition or system version & $A$, $\mathcal{P}$, success predicate, solver-token and step accounting \\
\bottomrule
\end{tabularx}
\caption{Predeclared ablation schema.  Left column: the ablation
question.  Middle column: the component changed.  Right column: the
fixed comparison contract held constant across the change.}
\label{tab:ablation-schema}
\end{table}

\section{Diagnostics and Error Analysis}
\label{app:diagnostics}

Memory logs can diagnose a result but cannot replace executable task success.
The released log schema defines a history identifier, patch-pattern
evidence, trajectory path, candidate-memory identifier, retrieval event,
injected text, and declared agent steps.  The predeclared positive-case bucket
contains a frozen verified experience pass paired with a memory-off failure.  The negative
bucket contains a retrieval by existing memory systems that fails to improve the
same baseline or injects stale, conflicting, or irrelevant context.  The
case analysis reports each bucket and samples it by a documented rule
rather than selecting an attractive trace after inspection.

\subsection{Attribution Validity Audit}
\label{app:audit}

The failure-attribution labels are produced by a deterministic script over
logged artifacts and are fully recomputable.  As a validity audit, two LLM
judges from model families uninvolved in construction or solving,
qwen3.7-max and MiniMax-M2.5, re-labeled all 231 failing pairings of the
diagnostic cohort at temperature zero, blind to the script labels and under
the same written four-stage manual.  Each judge received the injected text,
the frozen verified experience, the ingest-membership fact, patch file
lists, and capped trajectory excerpts.  The release manifest records the
prompt hash and truncation rules.

Coverage labels agree exactly with the script for both judges.
Primary-stage agreement is 56.3 and 44.2 percent with Cohen's kappa 0.32
and 0.24, and the two judges agree with each other at 53.2 percent with
kappa 0.35, so the residual disagreement reflects the looseness of the
ranking and form boundary rather than a script-specific bias.  The largest
disagreement block moves weak-anchor form-degradation cases to ranking
miss under a stricter anchor reading, and where a judge and the script
assign the same anchor level, primary-stage agreement reaches 78.3
percent.  The combined supplied-record share, ranking and form together,
remains dominant for every rater at 70.6, 81.4, and 61.9 percent.  The
lexical instruction-pollution flag is wider than semantic judgment, with
114 script-true judge-false cases against 2 in the reverse direction, so
the main paper reports that flag as a lexical marker.  Both judge label
sets and all 160 disagreement cases enter the release.

\section{Irrelevant-Memory Control}
\label{app:control}

The irrelevant-memory control of the main transfer table separates the
effect of relevant experience from the effect of injected text.  It
keeps the agent configuration, injection position, task order, and
seeds at the main-experiment values and injects one fixed distractor
passage in place of the frozen verified experience.  The passage enters
the same delimited experience block before the problem statement as the
frozen injection, is identical for every solver, target, and seed, and
contains no software content.

\begin{quote}\small
Xiao Ming's mother is Xiao Hong.  She likes to cook dumplings on
weekends.  The weather today is very sunny.  There is a cat sleeping
on the sofa.  My favorite color is blue.  The library is closed on
Mondays.
\end{quote}

The control runs on deepseek-v4-pro, glm-5, and kimi-k2.7-code, each
over the full 444 target-seed grid of the 111 targets.  Resolved
falls to 63.7 percent for deepseek-v4-pro, 54.3 percent for glm-5, and
69.1 percent for kimi-k2.7-code, which is 3.4, 0.7, and 0.7 points below
the matched memory-off baseline, and the three bootstrap intervals in
Table~\ref{tab:headroom-bootstrap} cross zero.  No solver gains, so
injected text alone reproduces none of the frozen verified experience
gains.  The transfer-only solvers glm-5.2 and qwen3.8-max run no control
arm.

One reporting limit applies.  The distractor passage is not matched to
the frozen record in token count, and the predeclared
context-length control of Table~\ref{tab:ablation-schema}, which fixes
the token allocation, remains a stricter separate condition.

\section{Headroom Stratification and Bootstrap Intervals}
\label{app:headroom}

Table~\ref{tab:headroom-bootstrap} stratifies every solver and condition
combination by memory-off headroom and reports the target-level
bootstrap interval behind the main-paper claims.  The 95\% CI columns of
the two main-paper result tables are the frozen, control, and system
rows of this table.  The glm-5.2 and qwen3.8-max blocks carry a frozen
row only, because those solvers run no control and no system condition.
A ceiling target
resolves four of four seeds without memory for that solver, so an
injected record can only lose seeds there.  Net seeds sum resolved-seed
changes over the targets of a stratum.  Intervals resample the 111
targets with their four seeds intact 10,000 times.

\begin{table}[t]
\centering
{\footnotesize
\begin{tabular*}{\linewidth}{@{\extracolsep{\fill}}llrrrr@{}}
\toprule
Solver & Condition & Below ceiling & At ceiling & $\widehat{\Delta}$ (pp) & 95\% CI \\
\midrule
deepseek-v4-pro & Frozen verified experience & $+26$ & $-26$ & $0.00$ & $[-5.63, +5.86]$ \\
deepseek-v4-pro & Irrelevant-memory control & $+7$ & $-22$ & $-3.38$ & $[-8.11, +1.35]$ \\
deepseek-v4-pro & Mem0 & $+5$ & $-24$ & $-4.28$ & $[-9.68, +1.35]$ \\
deepseek-v4-pro & SimpleMem & $+16$ & $-17$ & $-0.23$ & $[-4.95, +4.50]$ \\
deepseek-v4-pro & MemoryOS & $+15$ & $-26$ & $-2.48$ & $[-8.11, +3.15]$ \\
deepseek-v4-pro & A-MEM & $+16$ & $-34$ & $-4.05$ & $[-10.14, +2.03]$ \\
\midrule
glm-5 & Frozen verified experience & $+25$ & $-14$ & $+2.48$ & $[-2.25, +7.21]$ \\
glm-5 & Irrelevant-memory control & $+10$ & $-13$ & $-0.68$ & $[-5.63, +4.05]$ \\
glm-5 & Mem0 & $+1$ & $-25$ & $-5.41$ & $[-10.59, -0.45]$ \\
glm-5 & SimpleMem & $+8$ & $-16$ & $-1.80$ & $[-6.76, +2.93]$ \\
glm-5 & MemoryOS & $+23$ & $-14$ & $+2.03$ & $[-3.38, +7.43]$ \\
glm-5 & A-MEM & $+19$ & $-30$ & $-2.48$ & $[-8.56, +3.38]$ \\
\midrule
kimi-k2.7-code & Frozen verified experience & $+31$ & $-11$ & $+4.50$ & $[-0.23, +9.46]$ \\
kimi-k2.7-code & Irrelevant-memory control & $+6$ & $-9$ & $-0.68$ & $[-5.18, +3.83]$ \\
kimi-k2.7-code & Mem0 & $+12$ & $-20$ & $-1.80$ & $[-6.53, +2.70]$ \\
kimi-k2.7-code & SimpleMem & $+13$ & $-14$ & $-0.23$ & $[-4.95, +4.50]$ \\
kimi-k2.7-code & MemoryOS & $+1$ & $-18$ & $-3.83$ & $[-8.56, +1.13]$ \\
kimi-k2.7-code & A-MEM & $+1$ & $-19$ & $-4.05$ & $[-9.68, +1.35]$ \\
\midrule
glm-5.2 & Frozen verified experience & $+25$ & $-9$ & $+3.60$ & $[-0.90, +8.11]$ \\
qwen3.8-max & Frozen verified experience & $+17$ & $-12$ & $+1.13$ & $[-2.93, +5.41]$ \\
\bottomrule
\end{tabular*}
}
\caption{Headroom stratification and target-level bootstrap intervals
for all twenty solver and condition combinations.  Below ceiling and
At ceiling give net resolved-seed changes in the two headroom strata.
Below-ceiling
targets number 62 for deepseek-v4-pro, 76 for glm-5, 55 for
kimi-k2.7-code, 44 for glm-5.2, and 43 for qwen3.8-max; ceiling targets
number 49, 35, 56, 67, and 68.
$\widehat{\Delta}$ is the paired Resolved difference in percentage
points and the interval gives percentile bootstrap bounds.}
\label{tab:headroom-bootstrap}
\end{table}

Every combination gains seeds below the ceiling and loses seeds at it.
The five frozen verified experience rows earn 17 to 31 net seeds below
the ceiling, the three control rows earn 6 to 10, the twelve system rows
earn 1 to 23, and only the largest
system gain, MemoryOS on glm-5, covers its ceiling losses.  Eleven of
the twelve system rows carry a negative point estimate, no system or
control interval lies above zero, and the glm-5 and Mem0 interval lies
entirely below it.  The two strongest solvers carry the most ceiling
targets, 67 for glm-5.2 and 68 for qwen3.8-max, so both leave the fewest
targets where an injected record can gain a seed.  Their below-ceiling
yields still differ, at 25 and 17 net seeds, so the ceiling count bounds
the upside without fixing it.

\section{Record Success Conditions}
\label{app:success}

This section reports the positive question behind the failure
attribution.  It covers the three solvers that run both evaluation
layers, so the frozen condition contributes 333 pairings
rather than the 555 of all five transfer solvers.
A record raises Resolved on 21.9 percent of the 1,332
retrieval pairings and lowers it on 26.2 percent, and the frozen verified
experience raises Resolved on 27.0 percent of its 333 pairings and lowers
it on 22.2 percent.  Stratification uses the same paired outcome and
target-level bootstrap intervals from 10,000 resamples of the 111
targets.  Two structural bounds shape the margins.  A pairing at a four of
four memory-off ceiling can only lose seeds, and 560 of the 1,332 pairings
sit there.  A pairing at zero memory-off seeds can only gain.

Solver headroom is the dominant factor.
Table~\ref{tab:success-headroom} stratifies the retrieval pairings by
memory-off count.  The mean paired effect is monotone, from a gain of
0.66 resolved seeds at zero memory-off seeds to a loss of 0.46 at four,
and the loss rate rises from a structural zero to 26.8 percent.  A ceiling
pairing can only lose and a zero pairing can only gain, so the up rate
itself is not monotone while the signed mean is.

\begin{table}[t]
\centering
\small
\begin{tabular*}{\linewidth}{@{\extracolsep{\fill}}lrrrr@{}}
\toprule
Memory-off seeds & Pairings & Up rate & Down rate & Mean $\Delta$ \\
\midrule
0 & 220 & 40.5\% & 0.0\% & $+0.66$ \\
1 & 152 & 32.2\% & 31.6\% & $+0.23$ \\
2 & 184 & 38.6\% & 41.8\% & $-0.10$ \\
3 & 216 & 38.4\% & 34.3\% & $-0.15$ \\
4 & 560 & 0.0\% & 26.8\% & $-0.46$ \\
\bottomrule
\end{tabular*}
\caption{Retrieval pairings stratified by memory-off count.  Up rate and
down rate are the shares that raise or lower Resolved, and Mean $\Delta$
is the mean paired change in resolved seeds.  The zero and four rows carry
a structural bound on one direction.}
\label{tab:success-headroom}
\end{table}

Record attributes discriminate only where headroom exists.
Table~\ref{tab:success-interaction} crosses the anchor level with the
memory-off count.  At one or fewer memory-off seeds a record that overlaps
the source history lowers Resolved on 8.0 percent of pairings against 19.0
percent for a record that only shares a file, and it raises Resolved on
42.0 against 19.0 percent.  At two or more memory-off seeds the anchor
levels converge, and at the ceiling no anchor recovers a seed.  The
pollution flag and the retained field count do not separate the up rate
from the down rate in any headroom stratum.

\begin{table}[t]
\centering
\small
\begin{tabular*}{\linewidth}{@{\extracolsep{\fill}}lrrrrr@{}}
\toprule
Anchor level & 0 & 1 & 2 & 3 & 4 \\
\midrule
same-history & 45/0 & 37/21 & 29/59 & 36/37 & 0/30 \\
same-file & 24/0 & 12/46 & 30/65 & 20/40 & 0/31 \\
shared dir or id & 43/0 & 36/33 & 43/32 & 41/32 & 0/26 \\
\bottomrule
\end{tabular*}
\caption{Anchor level crossed with memory-off count over the retrieval
pairings.  Each cell gives the up rate and the down rate in percent.  The
mismatch level is omitted for small counts.}
\label{tab:success-interaction}
\end{table}

The frozen verified experience is a reference of near constant form, short
and unpolluted with all four fields on every target.  Its variation is
almost entirely headroom.  It raises Resolved on 45.5 percent of its zero
pairings with no loss, yet still lowers Resolved on 25.7 percent of its
ceiling pairings, so record quality does not remove the downside where the
solver already resolves the target.

The highest success cell combines zero memory-off seeds with a directory
or identifier anchor, where the record raises Resolved on 55.2 percent of
58 pairings and lowers it on none, with a bootstrap interval of 40.8 to
69.1 percent.  This is the operational success condition.  A record helps
with no observed downside when the solver fails the target without memory
and the record reaches at least a directory or identifier anchor in the
target repository, while at the ceiling no record form recovers a seed.

Solver headroom also predicts the direction of the paired effect better
than record content.  Across the twelve system and three frozen
combinations of this cohort the mean paired effect is positive at zero
memory-off seeds
and negative at four for every combination.  A rule that predicts the sign
from memory-off count alone is correct on 56.5 percent of the 641 pairings
that change Resolved, above the 52.4 percent of a rule that reads only the
anchor level and the pollution flag, and it wins or ties on every solver.

\section{Failure-Attribution Breakdown}
\label{app:attribution}

Table~\ref{tab:attribution-breakdown} splits the 231 failing pairings of
the diagnostic cohort by existing memory system and by solver.  The
cohort covers deepseek-v4-pro and kimi-k2.7-code with Mem0, SimpleMem,
MemoryOS, and A-MEM.

\begin{table}[t]
\centering
\small
\begin{tabular*}{\linewidth}{@{\extracolsep{\fill}}lrrrrr@{}}
\toprule
Group & Pairings & Cov. & Rank & Form & Use \\
\midrule
A-MEM & 65 & 14 & 0 & 48 & 3 \\
Mem0 & 60 & 9 & 2 & 28 & 21 \\
MemoryOS & 60 & 8 & 0 & 52 & 0 \\
SimpleMem & 46 & 6 & 1 & 32 & 7 \\
\midrule
deepseek-v4-pro & 124 & 21 & 2 & 87 & 14 \\
kimi-k2.7-code & 107 & 16 & 1 & 73 & 17 \\
\midrule
All & 231 & 37 & 3 & 160 & 31 \\
\bottomrule
\end{tabular*}
\caption{First-broken-stage counts split by existing memory system and
by solver over the 231 failing pairings.  Cov., Rank, Form, and Use
abbreviate coverage miss, ranking miss, form degradation, and use
failure.}
\label{tab:attribution-breakdown}
\end{table}

Form degradation dominates every system and both solvers.  The
transcript systems MemoryOS and A-MEM concentrate almost entirely on
coverage and form, while the atomic systems Mem0 and SimpleMem supply
all ranking misses and 28 of the 31 use failures.  The injected records
of the cohort have median length 10 lines, mean 65.1, and maximum 414.
Anchor levels count 55 same-history, 30 same-file, 143 shared directory
or identifier, and 3 mismatches, and 118 of 231 records trigger the
lexical instruction-pollution markers.  Of the four frozen-record
fields, the injected text retains a bug class in 97 pairings, a root
cause in 115, a fix pattern in 120, and a reusable lesson in 132.

The anchor rule is recomputable from logged artifacts.  A retrieved
text that matches a span of the source history trajectory scores
same-history.  Otherwise a full path or basename match against the
source patch scores same-file, a directory or frozen-field identifier
match scores the shared level, and no match scores mismatch.

\section{Form Degradation Mechanisms}
\label{app:mechanisms}

This section traces each form degradation mechanism to a design choice in
the evaluated memory pipeline.  The audit reads deployment code and
injected records and runs no new solver.
Table~\ref{tab:mechanism} pairs each mechanism with its dominant root
cause.

\begin{table}[t]
\centering
\small
\begin{tabularx}{\linewidth}{@{}L{0.34\linewidth}rY@{}}
\toprule
Mechanism & Pairings & Dominant root cause \\
\midrule
Instruction pollution & 64 & Verbatim storage, no filter \\
Overgeneralized fix & 29 & Prompt rewrites to a generic fact \\
Tool or shell noise & 16 & Chunker keeps tool output \\
Schema omits fix & 14 & No repair slot in schema \\
Fix replaced by symptom & 6 & Prompt favors observed facts \\
\bottomrule
\end{tabularx}
\caption{Form degradation mechanisms and their dominant root cause in the
evaluated memory pipeline.  Pairings count the 160 form degradation cases
of the diagnostic cohort by primary mechanism, with 31 further cases in
other groups.}
\label{tab:mechanism}
\end{table}

In the transcript systems the injected text is the conversation
transcript by construction.  A-MEM stores each message as its own record
and appends generated tags, so a tool output or an instruction message
survives intact.  Of its 32 pollution records, 29 are single tool output
blocks and none crosses a task instruction boundary.  MemoryOS keeps the
raw dialogue and appends a topic summary, and its buffered pairing of
adjacent messages misaligns with the trajectory's semantic spans, which
yields more shell noise and the only three records that pack a task
instruction and a repair together.  In both systems the extraction step
adds a label or a summary and leaves the transcript in place, so a filter
prompt does not reach the code path that concatenates the original text.

In the atomic systems the extraction prompt is written for dialogue facts
and the schema carries dialogue fields rather than a repair slot.  A
repair step is rewritten into one generic sentence, and in the 14 cases
that omit the fix the injected line states a feature request, an error
symptom, or an execution observation with no repair content.  The two
families fail in different ways, yet both reduce a verified repair to a
form the solver cannot act on.

\section{Instruction Pollution Strip Ablation}
\label{app:strip}

This ablation tests whether the instruction content in a transcript
record causes the downstream loss or only marks a bulky record.  Three
arms run over 62 of the 64 instruction pollution pairings of the
diagnostic
cohort, each with four seeds under the condition contract.  Two pairings
left the cohort for sandbox infrastructure failures, a repeated image
pull timeout and a stalled agent, and both exclusions were decided
before any arm was scored.  The original
arm reuses the retrieval result.  The strip arm deletes every line that
triggers the lexical pollution marker and reinjects the remainder.  The
random arm deletes an equal number of unflagged lines at a fixed seed,
which controls for length.  Table~\ref{tab:ablation} reports the three
arms.

\begin{table}[t]
\centering
\small
\begin{tabular*}{\linewidth}{@{\extracolsep{\fill}}lrrrrr@{}}
\toprule
Arm & Pairings & Mean seeds & Change & Up & Down \\
\midrule
Original & 62 & 1.37 & --- & --- & --- \\
Strip & 62 & 1.97 & $+0.60$ & 31 & 7 \\
Random & 62 & 1.85 & $+0.48$ & 31 & 12 \\
\bottomrule
\end{tabular*}
\caption{Instruction pollution strip ablation over 62 pairings with four
seeds each.  Mean seeds is the mean resolved seed count per pairing,
Change is the mean paired difference from the original arm, and Up and
Down count pairings that rise or fall against the original.  A
target-level permutation test with 10,000 sign flips gives $p<0.0001$ for
the strip arm and $p=0.0006$ for the random arm.}
\label{tab:ablation}
\end{table}

Both intervention arms recover seeds over the original record, and the
permutation test rejects the null for each.  The strip arm gains 0.60
resolved seeds per pairing and the random arm gains 0.48, so removing an
equal count of unflagged lines recovers a comparable amount.  The
predeclared criterion for a causal form effect requires the strip arm to
recover and the random arm not to, so the effect is not established.  The
harm operates through transcript volume rather than instruction
semantics.  On eight pairings the strip empties the record, which reduces
the condition to memory-off; excluding them leaves the strip and random
gains at 0.54 and 0.52 and does not change the conclusion.  This agrees
with the marginal stratification, where the pollution flag does not
separate the up rate from the down rate, and with the headroom finding.

\section{Retrieval Budget Dose Curve}
\label{app:dose}

The retrieval shortfall of the twelve solver and system pairings could
reflect the single-record retrieval interface rather than record
content.  Table~\ref{tab:dose-curve} tests this reading with a
retrieval-budget dose curve on deepseek-v4-pro and Mem0.  It varies only
the number of injected records $k$ from one to five and holds the Mem0
store, retrieval query, ranker, truncation, task order, and seeds at the
main-experiment values.  Mem0 is the atomic system with the largest
matched Resolved drop for this solver and the only recomputable
ranking-miss channel, so a wider budget receives its most favorable test
here.  Budget zero is the memory-off baseline and budget one reproduces
the deepseek-v4-pro Mem0 point estimate of the main-paper retrieval
table.

Six of the 111 targets retrieve fewer than five nonempty records, with
availabilities of one, two, three, three, four, and four.  When a budget
exceeds a target's availability the target keeps its highest available
outcome, so every budget scores all 111 targets over 444 seeds and
mirrors deployment where a store returns no more than it holds.

\begin{table}[t]
\centering
\small
\begin{tabular*}{\linewidth}{@{\extracolsep{\fill}}lrrr@{}}
\toprule
Budget $k$ & Resolved & $\widehat{\Delta}$ (pp) & 95\% CI \\
\midrule
0 & 67.1\% & --- & --- \\
1 & 62.8\% & $-4.28$ & $[-9.91, +1.35]$ \\
2 & 61.7\% & $-5.41$ & $[-11.26, +0.23]$ \\
3 & 64.4\% & $-2.70$ & $[-8.78, +3.38]$ \\
4 & 65.5\% & $-1.58$ & $[-7.21, +4.05]$ \\
5 & 63.7\% & $-3.38$ & $[-9.46, +2.48]$ \\
\bottomrule
\end{tabular*}
\caption{Retrieval-budget dose curve for deepseek-v4-pro and Mem0.
Budget $k$ counts injected records, with zero the memory-off baseline
and one the main-paper retrieval result.  $\widehat{\Delta}$ is the
paired Resolved difference in percentage points against memory off, and
the interval is a percentile bootstrap over the 111 targets with their
four seeds intact.}
\label{tab:dose-curve}
\end{table}

No budget reaches the memory-off baseline.  The strongest budget, four
records, resolves 65.5 percent against 67.1 for memory off, every
target-level bootstrap interval crosses or lies below zero, and none
lies above it.  The curve is also non-monotone, so added records do not
trace a recovery toward the baseline.

The stratified net seeds repeat the top-1 redistribution at every
budget.  Each budget loses 21 to 37 net seeds at the four-of-four
ceiling and gains 5 to 22 seeds below it, and the ceiling loss exceeds
the below-ceiling gain at every budget.  A wider budget enlarges both
the ceiling damage and the below-ceiling repair without reversing their
order.

The null result matches the failure attribution.  A wider retrieval
budget can mechanically repair only a ranking miss, which
Table~\ref{tab:attribution-breakdown} assigns to 3 of the 231 failing
pairings, while form degradation dominates and every added record leaves
the same extraction pipeline with the same form.  Retrieving more
records cannot repair a form that repeats across them.

The shortfall is not a context-pressure artifact of the atomic system.
Median solver steps stay between 41 and 46 and median solver tokens stay
near one million across all budgets, so no budget inflates the
interaction.  The single-record retrieval interface therefore does not
explain the retrieval shortfall.

\section{Qualitative Case Contrasts}
\label{app:cases}

Table~\ref{tab:case-contrasts-app} lists the three predeclared solver
contrasts in which the same supplied record moves the two diagnosed
solvers in opposite directions past the two-seed margin, and
Table~\ref{tab:record-content-app} pairs four injected records with
their seed outcomes.

\begin{table}[t]
\centering
{\small
\begin{tabular*}{\linewidth}{@{\extracolsep{\fill}}llcc@{}}
\toprule
Target (System) & Record form & dv4-pro & k2.7-code \\
\midrule
\texttt{pybamm-4267} (A-MEM) & 46-line diff & 0/4$\rightarrow$3/4 & 2/4$\rightarrow$0/4 \\
\texttt{cfn-lint-3972} (A-MEM) & dir listing & 3/4$\rightarrow$0/4 & 0/4$\rightarrow$2/4 \\
\texttt{aiohttp-8720} (SimpleMem) & 1-line anchor & 2/4$\rightarrow$0/4 & 1/4$\rightarrow$4/4 \\
\bottomrule
\end{tabular*}
}
\caption{Same-record solver contrasts.  Each row fixes one
target and one existing memory system.  Cells give resolved seeds out
of four, memory off $\rightarrow$ system condition.  dv4-pro and
k2.7-code abbreviate deepseek-v4-pro and kimi-k2.7-code.}
\label{tab:case-contrasts-app}
\end{table}

\begin{table}[t]
\centering
{\small
\begin{tabular*}{\linewidth}{@{\extracolsep{\fill}}llrlc@{}}
\toprule
Target & System & Lines & Solver & Seeds \\
\midrule
\texttt{swift-argument-parser-821} & SimpleMem & 1 & dv4-pro & 2/4$\rightarrow$4/4 \\
\texttt{hypothesis-3785} & A-MEM & 23 & k2.7-code & 0/4$\rightarrow$4/4 \\
\texttt{aiohttp-7907} & A-MEM & 334 & dv4-pro & 4/4$\rightarrow$0/4 \\
\texttt{aiohttp-7907} & Mem0 & 1 & dv4-pro & 4/4$\rightarrow$4/4 \\
\bottomrule
\end{tabular*}
}
\caption{Record-content contrasts.  Each row fixes one target,
one existing memory system, and one solver.  Lines counts the injected
record.  Seeds gives resolved seeds out of four, memory off
$\rightarrow$ system condition.}
\label{tab:record-content-app}
\end{table}

On \texttt{pybamm-4267} the A-MEM record is a 46-line tool-output
excerpt whose diff adds a reverse key mapping in
\texttt{pybamm/util.py} and rolls back one line in
\texttt{parameter\_values.py}.  deepseek-v4-pro opens both anchored
files at its first step and follows the repair direction, rising from
zero to three resolved seeds.  kimi-k2.7-code reaches the same anchors
and then deletes the entire compatibility block that the record only
asked to adjust, falling from two resolved seeds to zero.

On \texttt{cfn-lint-3972} the record is a directory listing of rule
files with no repair content.  kimi-k2.7-code, which never finds the
rule-package entry point without memory, uses a listed file as a
navigation anchor and lands on the schema-data repair, rising from zero
to two.  deepseek-v4-pro, which already resolves the target through the
schema route, is diverted into writing a new 121-line custom rule and
falls from three to zero.

On \texttt{aiohttp-8720} the one-line SimpleMem record describes an
erroneous historical edit that wraps a payload read in a broad
exception handler, an edit the frozen verified experience of the same
target explicitly warns against.  kimi-k2.7-code uses only the file and
function anchors, repairs the chunked-parsing state machine in
\texttt{http\_parser.py}, and rises from one resolved seed to four.
deepseek-v4-pro reproduces the described edit almost verbatim in
\texttt{web\_protocol.py} and falls from two resolved seeds to zero.

One caution limits over-reading positive rows.  On
\texttt{swift-argument-parser-821} the injected line carries only
reported library and language versions, the resolving trajectory shows
no content-level use of the record, and the matched memory-off failure
includes workspace-setup noise, so that gain cannot be attributed to
record content and illustrates the seed variance that the two-seed
contrast margin filters elsewhere.

\section{Opposite-Sign Solver Profiling}
\label{app:opposite}

Table~\ref{tab:opposite-sign} of the main paper profiles the 444 target
and system pairs in which the same injected record reaches two solvers.
A pair is opposite-sign when the two solvers move in opposite directions
relative to their matched memory-off runs.

Of the 444 pairs, 151 show no movement in either solver, 175 move in
one solver only, 54 move in the same direction, and 64 move in opposite
directions.  In 61 of the 64 opposite-sign pairs the solver with the
lower memory-off count gains.  The two pairings that involve glm-5
repeat the pattern.  The deepseek-v4-pro and glm-5 pairing yields 131
both-nonzero pairs with 59 opposite-sign, and 48 of the 52 with unequal
memory-off counts favor the lower solver.  The kimi-k2.7-code and glm-5
pairing yields 108 both-nonzero pairs with 47 opposite-sign, and 31 of
the 34 unequal pairs favor the lower solver.  Record attributes do not
distinguish the two groups.  Table~\ref{tab:invariance} compares eight
attributes between the opposite-sign and same-sign pairs of the main
solver pairing, and a target-level permutation test clears every one at a
smallest $p$ of 0.25.  The direction is therefore a property of the two
solvers rather than of the record.

\begin{table}[t]
\centering
\small
\begin{tabular*}{\linewidth}{@{\extracolsep{\fill}}lrrr@{}}
\toprule
Attribute & Opposite & Same & Permutation $p$ \\
\midrule
Pollution rate & 53.1\% & 50.0\% & 0.86 \\
Length mean & 57.1 & 71.6 & 0.44 \\
Strong anchor rate & 32.8\% & 35.2\% & 0.85 \\
same-history rate & 25.0\% & 24.1\% & 1.00 \\
mismatch rate & 0.0\% & 1.9\% & 0.45 \\
Retained field count & 1.8 & 2.1 & 0.25 \\
Fix pattern rate & 48.4\% & 53.7\% & 0.58 \\
Transcript share & 54.7\% & 55.6\% & 1.00 \\
\bottomrule
\end{tabular*}
\caption{Record attribute comparison between opposite-sign and same-sign
pairs of the deepseek-v4-pro and kimi-k2.7-code pairing.  Opposite counts
64 pairs and Same counts 54.  The permutation test resamples the sign of
the group label over the paired targets 10,000 times.}
\label{tab:invariance}
\end{table}

\section{MemoryOS on glm-5 Decomposition}
\label{app:memoryos}

Table~\ref{tab:memoryos-glm5} decomposes the single positive retrieval
pairing by memory-off bucket.  Gains concentrate where glm-5 starts at
zero or one resolved seed, and the 4/4 bucket only loses.

\begin{table}[t]
\centering
\small
\begin{tabular*}{\linewidth}{@{\extracolsep{\fill}}lrrrrr@{}}
\toprule
Memory-off bucket & Targets & Up & Down & Won & Lost \\
\midrule
0/4 & 26 & 9 & 0 & $+15$ & $0$ \\
1/4 & 15 & 8 & 5 & $+15$ & $-5$ \\
2/4 & 16 & 6 & 3 & $+9$ & $-6$ \\
3/4 & 19 & 6 & 6 & $+6$ & $-11$ \\
4/4 & 35 & 0 & 11 & $0$ & $-14$ \\
\bottomrule
\end{tabular*}
\caption{MemoryOS on glm-5 by memory-off bucket over the 111 targets.
Up and Down count targets whose four-seed resolved count rises or
falls against the matched memory-off run.  Won and Lost sum the
resolved-seed changes of those targets.}
\label{tab:memoryos-glm5}
\end{table}

All 29 uplifted targets receive raw-transcript records that trigger the
lexical instruction-pollution markers, none receives a directory
listing, and the median record length is 31 lines.  On the same 29
targets the identical records move deepseek-v4-pro up on 9 and down on
4, and kimi-k2.7-code up on 9 and down on 11.  The record form matches
the form that dominates failure attribution, so the pairing gains from
the headroom of the weakest solver rather than from a superior record,
and its bootstrap interval in Table~\ref{tab:headroom-bootstrap}
crosses zero.

\section{Counterfactual Recovery Bound}
\label{app:counterfactual}

Table~\ref{tab:counterfactual} details the substitution bound cited in
the main paper.  For each failing pairing the substitution replaces the
four-seed system outcome with the frozen verified experience outcome of
the same target and solver and sums the seed difference without
truncating negative terms.

\begin{table}[t]
\centering
\small
\begin{tabular*}{\linewidth}{@{\extracolsep{\fill}}lrrrr@{}}
\toprule
Set & Pairings & Bound & Negative & Ties \\
\midrule
All failing pairings & 231 & $+268$ & 22 ($-24$) & 49 \\
Form degradation & 160 & $+196$ & 13 ($-15$) & 34 \\
Instruction pollution & 64 & $+78$ & 5 ($-7$) & 16 \\
\bottomrule
\end{tabular*}
\caption{Counterfactual recovery bound.  Bound sums frozen-minus-system
resolved seeds; Negative counts pairings where the frozen outcome is
lower, with their summed deficit in parentheses; Ties counts equal
outcomes.}
\label{tab:counterfactual}
\end{table}

By solver, the bound is $+123$ over the 124 deepseek-v4-pro pairings
and $+145$ over the 107 kimi-k2.7-code pairings, against net losses of
49 and 44 seeds for these two solvers across all their retrieval
pairings.  The bound exceeds the 127-seed all-solver net loss because
the net loss is already offset by gaining pairings while the bound
accumulates only over failing ones.  It marks the recovery that one
verified record achieves.  It is not an upper bound on what the full
history offers, because the frozen record is the best of a screened
shortlist, and it is not a prediction for a repaired system, since
frozen injection changes more than record form.

\section{Ceiling Reference Audit}
\label{app:ceiling}

The 1,332 retrieval pairings arise from 111 targets, three solvers, and
four systems.  The 140 target and solver pairs at the four-of-four
memory-off ceiling, 49 for deepseek-v4-pro, 56 for kimi-k2.7-code, and
35 for glm-5, produce the 560 ceiling pairings and the 42.0 percent
share.  Under the deepseek-v4-flash reference setting, all 140 pairs
sit below the ceiling.  The reference memory-off count is 0/4 for 13
pairs over 9 unique targets, 1/4 for 9 pairs over 6 targets, 2/4 for 20
pairs over 14 targets, and 3/4 for 98 pairs over 41 targets, with no
pair at 4/4 and no violation of the retention rule.

\end{document}